\documentclass[aps,prstab,twocolumn,groupedaddress]{revtex4-2}
\usepackage{graphicx}
\usepackage{amsmath}
\usepackage{xcolor}
\usepackage{url}
\begin{document}

%\preprint{}

\title{Detuning Properties of RF Phase Modulation in an Electron Storage Ring}

\author{A. Mochihashi}\email[Contact author: ]{akira.mochihashi@kit.edu}
\author{S. Maier} 
\author{E. Blomley}
\author{M. Schuh}
\author{E. Huttel}
\author{T. Boltz}
\altaffiliation{Present address: SLAC National Laboratory, Menlo Park, 94025, USA}
\author{B. Kehrer}
\author{A.-S. M{\"u}ller}
\affiliation{Karlsruhe Institute of Technology, Hermann-von-Helmholtz-Platz 1, 76344 Eggenstein-Leopoldshafen, Germany}

\author{D. Teytelman}
\affiliation{Dimtel, Inc., 2059 Camden Avenue, San Jose, California 95124, USA}
%\thanks{}
%\affiliation{}


\date{\today}

\begin{abstract}
In electron storage rings, it is possible to increase the bunch length by applying phase modulation to the radio frequency accelerating field by choosing appropriate parameters for the modulation. Such a bunch lengthening effect improves beam parameters such as beam lifetime, which can help us achieve better beam stability. It is well known that the modulation frequency around the double synchrotron frequency is effective in lengthening the bunch. The dependence of bunch lengthening on modulation frequency, the so-called detuning property, has a peak around the double synchrotron frequency with a frequency width that depends on the modulation amplitude. Nonlinear effects due to phase modulation lead to a peak frequency shift in a negative direction from the double synchrotron frequency, accompanied by an asymmetric peak shape. Beam current also affects the properties of the detuning condition of such bunch lengthening. We investigated the detuning property using a theoretical model and systematic measurements at the electron storage ring KARA, and we verified the qualitative agreement between the experiments and the theoretical model. The macro-particle simulations have confirmed the agreement. 
\end{abstract}

%\keywords{}

\maketitle

\section{\label{Introduction}Introduction} 
In high-energy electron storage rings dedicated to high-energy physics and synchrotron radiation experiments, it is essential to prepare stable beams for long-term, precise measurements and experiments. The beam lifetime is one of the parameters that characterize the beam stability and quality. It corresponds to the loss rate of the particles in the stored beam per unit of time. To compensate for beam loss, an operational scheme called top-up operation has now been widely applied to electron storage ring facilities. In this scheme, an additional electron beam is repeatedly injected from the injection accelerator into the storage ring to keep the electron beam current constant. However, the beam lifetime is still significant in the top-up operation because the lifetime determines the repetition rate of the beam injection, which relates to accelerator operational issues such as radiation protection, energy consumption, and operational stability.

 There are several processes by which the electrons in the beam are lost, and one of the dominant processes originates from a scattering process between electrons in the same electron bunch. In the process, a pair of electrons is scattered transversely via the betatron oscillation, and a part of the transverse momentum is converted to the longitudinal momentum. As a result, the scattered electrons are lost if the modified longitudinal momentum is beyond the momentum acceptance of the accelerator. This beam-loss process and the beam lifetime accordingly are called the Touschek effect and lifetime \cite{Touschek_1963}. The Touschek effect can be relaxed by reducing the electron density in the bunch because the beam loss comes from the scattering between electrons in the same bunch.

Elongating the longitudinal electron bunch distribution can mitigate the scattering process and serve as a countermeasure against the Touschek effect. In the Photon Factory electron storage ring at the High Energy Accelerator Research Organization (KEK), they experimented with bunch lengthening by applying a phase modulation to the radio-frequency (RF) accelerating voltage at double synchrotron frequency and observed an improvement in the beam lifetime \cite{SakanakaPRSTAB2000, SakanakaJJAP2001}. In this previous study, they also discussed the beam dynamics under phase modulation using a theoretical model and simulations based on macro-particle tracking. The experimental results and simulations have shown that the longitudinal distribution of the electrons tends to exhibit the quadrupole oscillation mode accompanied by bunch lengthening. In Figure 7 of reference \cite{SakanakaJJAP2001}, the authors discussed the frequency-detuning character of equilibrium oscillation amplitudes of electrons, including the nonlinear effect, using a theoretical model. In the Brazilian synchrotron light source, they experimented with suppressing the longitudinal coupled-bunch instability through RF phase modulation 
\cite{AbreuPRSTAB2006}. This previous study discussed the beam dynamics based on Hamiltonian formalism and the existence of two or three stable fixed points in the longitudinal phase space due to RF phase modulation. They also showed the change in the RF bucket structure for the phase modulation frequency, which implies a detuning property of bunch lengthening for the modulation frequency.

At the 2.5 GeV electron storage ring KARA (Karlsruhe Research Accelerator) \cite{IBPTwebsite} in Karlsruhe Institute of Technology (KIT), we have introduced the function of RF phase modulation into the low-level RF (LLRF) system of KARA, allowing us to change both the frequency and amplitude of the RF phase modulation quite easily. Because the maximum beam energy from the injection accelerator in KARA is 500 MeV, we ramp the beam energy up to 2.5 GeV with the storage ring during regular accelerator operation after the injection. Therefore, it is essential for KARA to improve the beam lifetime so that the experiments with the synchrotron radiation take place with enough beam stability. To introduce the RF phase modulation scheme into KARA accelerator operation, we have searched for the optimum modulation conditions to elongate the longitudinal bunch length by focusing on the modulation frequency and amplitude. In the search process, we have discussed the detuning conditions of bunch lengthening by scanning the modulation frequency around the double synchrotron frequency (2$f_s$). The frequency-detuning characteristics of the bunch elongation have shown a peak structure around the double synchrotron frequency.
To formulate the detuning properties, we have discussed a theoretical model that shows the dependence of the bunch length on the modulation frequency. The model indicates that the detuning curve should have a peak around the double synchrotron frequency, with a frequency shift in the negative direction accompanied by an asymmetric peak shape that corresponds to the experimental and simulation results. %\textcolor{blue}{The model showed the achievable bunch length and detuning frequency width. Both are relevant to optimizing phase modulation. These values are determined by phase modulation and accelerator operation parameters and align with experimental and simulated results.}

In the following part, we discuss the theoretical model, which describes the detuning character of the bunch length on the phase modulation parameters in section \ref{TheoreticalModel}. In section \ref{Experiment}, we discuss the bunch length measurement using the streak camera and the phase modulation under several different conditions. In section \ref{Simulation}, we discuss the simulation results for the longitudinal beam dynamics based on the macro-particle model under phase modulation.  Finally, we summarize our discussion in section \ref{Summary}. In this paper, unless otherwise specified, the bunch length refers to the root mean square values.

\section{\label{TheoreticalModel}Theoretical model}
In reference 
\cite{SakanakaPRSTAB2000}, they discuss the motion of the electron under the influence of RF phase modulation with a modulation frequency that is double the synchrotron oscillation frequency. In this section, we also discuss the motion of the electron in the same manner as \cite{SakanakaPRSTAB2000}, but by adding a detuning condition to the modulation frequency \cite{SakanakaJJAP2001}. We formulate the equation of motion in the case where the modulation frequency is slightly different from double the synchrotron oscillation frequency, and we derive the dependence of the electron's amplitude and the bunch length on the detuning condition. First, we discuss the electron's motion by solving its equation in the longitudinal direction using a linear approximation. In this approximation, the electron is affected by a linear restoring force that depends on time due to phase modulation. Next, we extend the discussion to a nonlinear approximation, in which the electron is affected by a nonlinear restoring force due to the large amplitude of the synchrotron oscillation. From these discussions, we derive a formula showing the dependence of bunch elongation on the difference in frequency between the double synchrotron frequency and the modulation frequency.

The equation of motion of the electron for the longitudinal direction can be written as \cite{SakanakaPRSTAB2000}
\begin{align}
\frac{d\tau}{dt} &= -\alpha_c \delta, \label{eq1_sec2} \\
\frac{d\delta}{dt} &= \frac{eV_c \cos(\phi_0 - \omega \tau + \phi_m)-U_0}{T_0 E_0} - 2\gamma_\epsilon \delta. 
\label{eq2_sec2}
\end{align}
Here $\tau$ and $\delta$ correspond to the time and relative energy difference between the electron and the synchronous particle under equilibrium conditions in the longitudinal phase space, $\alpha_c$ is the momentum compaction factor, $e$ is the unit charge, $V_c$ is the amplitude of the RF acceleration voltage, $\phi_0$ is the synchronous phase of the electron beam, $\omega$ is the angular frequency of the RF accelerating field, $U_0$ is the energy loss of the electron due to synchrotron radiation, $T_0$ is the revolution period of the electron, and $E_0$ is the electron beam energy. The second term on the right side of Eq. (\ref{eq2_sec2}) corresponds to the radiation damping term with the factor $\gamma_\epsilon$,
\begin{equation}
\gamma_\epsilon = \frac{1}{2T_0} \left( \frac{dU}{dE} \right)_{E_0}.
\label{eq3_sec2}
\end{equation}
Here $U$ is the synchrotron radiation loss, and $E$ is the beam energy. The phase $\phi_m$ represents the phase modulation term, and we now assume $\phi_m$ has the following modulation characteristics \cite{SakanakaPRSTAB2000, SakanakaJJAP2001}:
\begin{equation}
\phi_m = \phi_{m0} \cos \left( \left(2 \omega_s + \Delta \omega \right) t \right),
\label{eq4_sec2}
\end{equation}
where $\omega_s$ is the synchrotron angular frequency and given as $\omega_s^2=\frac{\alpha_c eV_c \omega \sin\phi_0}{T_0 E_0}$, $\phi_{m0}$ is the modulation amplitude, and $\Delta \omega$ corresponds to the angular detuning frequency of the phase modulation from $2 \omega_s$. If we assume that the modulation amplitude $\phi_{m0}$ is small, we can apply the following approximation for the cosine part in Eq. (\ref{eq2_sec2}) \cite{SakanakaPRSTAB2000}
\begin{equation}
\begin{aligned}
    &\cos \left(\phi_0+\phi_{m}-\omega\tau \right) \approx \\
    &\cos\phi_0 - \phi_{m}\sin\phi_0+\left(\sin\phi_0+\phi_{m}\cos\phi_0\right)\omega\tau.
    \label{eq5_sec2}
\end{aligned}
\end{equation}
This means that we treat the modulation amplitude $\phi_{m0}$ and the amplitude of the synchrotron oscillation $\tau$ as the infinitesimal values of the first order. 
By applying the linear approximation, we obtain the equation of motion for the electron performing synchrotron oscillation with phase modulation \cite{SakanakaPRSTAB2000}:
\begin{equation}
\begin{aligned}
\frac{d^2 \tau}{dt^2} + 2\gamma_\epsilon \frac{d \tau}{dt}
+& \omega_s^2 \left( 1+ \epsilon \cos (2\omega_s + \Delta \omega) t \right)\tau = \\
&\frac{\omega_s^2}{\omega}\phi_{m0}\cos\left(\left(2\omega_s+\Delta\omega\right)t\right),
\label{eq6_sec2}
\end{aligned}
\end{equation}
where $\epsilon = \phi_{m0} \cot \phi_0$. The left side of Eq. (\ref{eq6_sec2}) shows the harmonic oscillator with phase modulation and resistance proportional to the time derivative of the synchrotron oscillation amplitude. The right side shows the driving force to the harmonic oscillator with a driving frequency of $2\omega_s+\Delta\omega$, which is far from the resonant frequency $\omega_s$ of the harmonic oscillator and has less effect on the electron's motion. Therefore, we ignore the driving term in Eq. (\ref{eq6_sec2}) and obtain the following equation \cite{SakanakaPRSTAB2000}: 
\begin{equation}
\frac{d^2 \tau}{dt^2} + 2\gamma_\epsilon \frac{d \tau}{dt}
+ \omega_s^2 \left( 1+ \epsilon \cos (2\omega_s + \Delta \omega) t \right)\tau = 0. 
\label{eq7_sec2}
\end{equation}
To solve Eq. (\ref{eq7_sec2}), we assume the following solution for $\tau$:
\begin{equation}
\tau = a(t) \sin \omega_s t + b(t) \cos \omega_s t,
\label{eq8_sec2}
\end{equation}
with the following restriction \cite{book_oscillation}:
\begin{equation}
\dot{a}(t) \sin \omega_s t + \dot{b}(t) \cos \omega_s t = 0.
\label{eq9_sec2}
\end{equation}
By substituting Eq. (\ref{eq8_sec2}) and (\ref{eq9_sec2}) into Eq. (\ref{eq7_sec2}) we obtain the following equation:
%From Eq. (\ref{eq7_sec2}-\ref{eq9_sec2}), we obtain the following equation:
%
%
\begin{equation}
\begin{aligned}
    &\dot{a}(t)\cos\omega_s t - \dot{b}(t)\sin\omega_s t
    + 2\gamma_\epsilon \left( a(t)\cos\omega_s t - b(t)\sin\omega_s t \right) \\
    &+ \epsilon \omega_s \cos\left(2\omega_s t + \Delta\omega t \right)
    \left( a(t) \sin\omega_s t + b(t)\cos\omega_s t \right)=0. \label{eq10_sec2}
\end{aligned}
\end{equation}
From Eqs. (\ref{eq9_sec2}) and (\ref{eq10_sec2}), the time derivative of $a(t)$ and $b(t)$ can be derived. If we only consider the change in the amplitude functions $a(t)$ and $b(t)$ with slow variation compared to the synchrotron oscillation period, we obtain the following equations for $a(t)$ and $b(t)$:
%Multiplying $\cos\omega_s t$ and $\sin\omega_s t$ on Eq. (\ref{eq10_sec2}) and considering only the changes with slow variation compared to the synchrotron oscillation period, we derive the equations for the amplitude functions $a(t)$ and $b(t)$ in Eq. (\ref{eq8_sec2}):
%
%
\begin{equation}
\begin{aligned}
\dot{a}(t) = &a(t) \Bigl( 
    -\gamma_\epsilon 
    + \frac{\omega_s \epsilon}{4}\sin\Delta\omega t 
    \Bigr) \\
    -&b(t) \frac{\omega_s \epsilon}{4} \cos\Delta \omega t,
    \label{eq11_sec2}
\end{aligned}
\end{equation}
\begin{equation}
\begin{aligned}
\dot{b}(t) = 
    &-a(t) \frac{\omega_s \epsilon}{4} \cos\Delta\omega t \\
    &- b(t) \left( \gamma_\epsilon+ 
    \frac{\omega_s \epsilon}{4}\sin\Delta\omega t \right).
    \label{eq12_sec2}
\end{aligned}
\end{equation}
The amplitude functions $a(t)$ and $b(t)$ are proportional to $\exp(-\gamma_\epsilon t)$ due to the radiation damping. If we write $a(t)$ and $b(t)$ as $a(t)=a_1(t) \exp(-\gamma_\epsilon t)$ and $b(t)=b_1(t) \exp(-\gamma_\epsilon t)$, we can express Eq. (\ref{eq11_sec2}) and (\ref{eq12_sec2}) as
\begin{equation}
    \left(
    \begin{array}{c}
    \dot{a_1}(t) \\
    \dot{b_1}(t)
    \end{array}
    \right)
    = \alpha
    \left(
    \begin{array}{cc}
    \sin\Delta\omega t     & -\cos\Delta\omega t \\
    -\cos\Delta\omega t     & -\sin\Delta\omega t
    \end{array}
    \right)
    \left(
    \begin{array}{c}
    a_1(t) \\
    b_1(t)
    \end{array}
    \right),
\label{eq13_sec2}
\end{equation}
where $\alpha=\frac{\omega_s \epsilon}{4}$. 
To discuss the motion of the electron, we introduce the value $\xi^2(t)=a_1^2(t)+b_1^2(t)$ and consider the behavior of $\xi(t)$ under two different conditions, namely the far resonance condition, where $\Delta \omega$ has a large value and the near resonance condition, where $\Delta \omega$ is nearly zero.

Under the far resonance condition, the electron does not continuously receive energy from the external force, and the electron's oscillation amplitude changes over time without diverging. The time period for the change is given by $\Delta \omega$, and the value $\left<\xi^2 \right>$, which is the average value of $\xi^2(t)$ over the time $\frac{2\pi}{\Delta \omega}$, is written as
%\sout{From the solution of Eq. (\ref{eq13_sec2}), we introduce the oscillation amplitude without the radiation damping term $\xi^2(t)=a_1^2(t)+b_1^2(t)$ and its average over the time period $\frac{2\pi}{\Delta\omega}$. The averaged oscillation amplitude $\left<\xi^2\right>$ shows the dependence of the elongation of the bunch length on the phase modulation parameters and is written as}
%
%
\begin{equation}
    \left<\xi^2\right>=\left<a_1^2\right>+\left<b_1^2\right>
    \approx \xi_0^2 \left( 1+\left(\frac{\alpha}{\Delta \omega}\right)^2
    \right),    
    \label{eq14_sec2}
\end{equation}
%
%
%\sout{where $\xi_0$ is a constant and is related to the natural bunch length.}
where $\xi_0$ is the parameter related to the natural bunch length.
%\sout{The derivation from Eq. (\ref{eq13_sec2}) to Eq. (\ref{eq14_sec2}) is summarized in Appendix \ref{appendix1}. The oscillation amplitude can be approximately written as}
The bunch length under phase modulation $\sigma_\text{mod}$ is written as
\begin{equation}
\sigma_\text{mod}\approx \sigma_\tau \sqrt{1+\left(\frac{\alpha}{\Delta\omega}\right)^2},
    \label{eq15_sec2}
\end{equation}
where $\sigma_\tau$ is the natural bunch length. 
%The derivation from Eq. (\ref{eq13_sec2}) to Eq. (\ref{eq14_sec2}) and (\ref{eq15_sec2}) is summarized in Appendix \ref{appendix1}. 
The derivation from Eq. (\ref{eq13_sec2}) to (\ref{eq15_sec2}) is summarized in Appendix \ref{appendix1}. 
From Eq. (\ref{eq15_sec2}), the detuning frequency $\Delta \omega_\rho$ where the bunch length $\sigma_{\text{mod}}$ is equal to $\rho \sigma_\tau$ $(\rho>1)$ is given as
\begin{equation}
     \Delta \omega_\rho = \pm \frac{\alpha}{\sqrt{\rho^2-1}}.
     \label{eq16_sec2}
\end{equation}
Especially, when the bunch lengthening factor $\rho$ is equal to $\sqrt{2}$, the detuning frequency width $\Delta f_{\sqrt{2}}$ where the bunch length is larger than $\sqrt{2}\sigma_\tau$ is given by
 \begin{equation}
     \Delta f_{\sqrt{2}} = \frac{\alpha}{\pi},
     \label{eq17_2_sec2}
 \end{equation}
where $\Delta f_{\sqrt{2}}$ has the dimension of Hz.

Under the near resonance condition, the electron continuously receives energy from the external force, and the oscillation amplitude of the electron increases over time. In this case, the value $\xi^2(t)$ is written as
 \begin{equation}
     \xi^2(t) = \frac{\xi_0^2}{2} \left(F_+(t)e^{-\frac{2\alpha}{\Delta \omega}}+F_-(t)e^{\frac{2\alpha}{\Delta \omega}}\right),
     \label{eq18_sec2}
\end{equation}
where
\begin{equation}
    \begin{aligned}
    F_{+}(t) = \left(1\pm\alpha\left(1+\frac{2\alpha}{\Delta \omega}\right)t\right)^2,\\
    F_{-}(t) = \left(1\pm\alpha\left(1-\frac{2\alpha}{\Delta \omega}\right)t\right)^2.
    \label{eq19_sec2}
    \end{aligned}
\end{equation}
%
%
%The derivation from Eq. (\ref{eq13_sec2}) to Eq. (\ref{eq18_sec2}) and (\ref{eq19_sec2}) is summarized in Appendix \ref{appendix1}. 
The derivation from Eq. (\ref{eq13_sec2}) to (\ref{eq18_sec2}) is summarized in Appendix \ref{appendix1}. 
%The bunch length $\sigma_{\text{mod}}$ under the near resonance is given as
%
%
% \begin{equation}
%     \sigma_\text{mod}=\sigma_\tau \sqrt{
%     1+\frac{1}{2}\left( \left(F_+(t)-1\right)e^{-\frac{2\alpha}{\Delta \omega}} + \left(F_-(t)-1\right)e^{\frac{2\alpha}{\Delta \omega}}\right)}.
%     \label{eq20_sec2}
% \end{equation}
%
%
This means that the oscillation amplitude of the electron increases monotonically, and the bunch length diverges under the near resonance condition with the linear approximation.

To consider the motion of the electron under phase modulation precisely, we introduce nonlinear approximation by expanding the cosine part of the equation of motion shown in Eq. (\ref{eq2_sec2}) up to the third-order terms, namely \cite{SakanakaJJAP2001}
%\sout{The detuning frequency width $|\Delta\omega_\rho|$ where the oscillation amplitude becomes $\rho\xi_0$ $(\rho>1)$ is
%
%
% \begin{equation}
%     \frac{|\Delta\omega_{\rho}|}{\omega_s} = \frac{\phi_{m0}\cot\phi_0}{2\sqrt{\rho^2-1}}.
%     \label{eq16_sec2}
% \end{equation}
%
%
% The bunch length increases to $\rho$ times the natural bunch length in the detuning frequency region of $-|\Delta \omega_\rho| \sim |\Delta \omega_\rho|$. Especially, the detuning frequency width where the bunch length lengthens to $\sqrt{2}$ times the natural bunch length is written as
% %
% %
% \begin{equation}
%     \frac{2|\Delta\omega_{\sqrt{2}}|}{\omega_s} = \phi_{m0}\cot\phi_0.
%     \label{eq17_sec2}
% \end{equation}
% %
% %
% The Eq. (\ref{eq16_sec2}) and (\ref{eq17_sec2}) mean that the width of the detuning curve depends on the modulation amplitude and the synchronous phase. The detuning curve (\ref{eq15_sec2}) has a peak at $\Delta \omega = 0$ with a symmetric form for the positive and negative detuning frequency sides. \\
% To consider the detuning character with non-linear approximations, we expand the cosine part of the equation of motion up to the third-order terms:
% }
%
%
\begin{equation}
\begin{aligned}
    &\cos(\phi_0+\phi_m-\omega\tau)\approx
    \cos\phi_0-\phi_m\sin\phi_0 \\
    &+\left(\sin\phi_0+\phi_m\cos\phi_0\right)\omega\tau
     +\frac{1}{2}\left(\phi_m\sin\phi_0-\cos\phi_0\right)\left(\omega\tau\right)^2 \\
    &-\frac{1}{6}\sin\phi_0\left(\omega\tau\right)^3.
    \label{eq21_sec2}
\end{aligned}
\end{equation}
We obtain the following equation of motion for the non-linear approximation with the cosine part and the approximations, which are the same as in the linear approximation:
\begin{equation}
\begin{aligned}
    &\frac{d^2\tau}{dt^2}+2\gamma_\epsilon\frac{d\tau}{dt} \\
    &+\Omega_s^2\left(
    1+\epsilon\sp{\prime}\cos\left(2\Omega_s+\Delta\omega+\frac{1}{8}\Omega_s\omega^2\Xi^2(t)\right)t\right)\tau=0,
    \label{eq22_sec2}
\end{aligned}
\end{equation}
where
\begin{align}
    \Xi(t) &= \sqrt{a^2(t)+b^2(t)}, 
    \label{eq23_sec2}\\
    \Omega_s^2&=\omega_s^2\left(1-\frac{1}{8}\omega^2\Xi^2(t)\right),
    \label{eq24_sec2}\\
    \epsilon\sp{\prime}&=\frac{\phi_{m0}}{1-\frac{1}{8}\omega^2\Xi^2(t)}\cot\phi_0.
    \label{eq25_sec2}
\end{align}
The derivation of Eq. (\ref{eq22_sec2}) is summarized in Appendix \ref{appendix2}. If we assume that the change in the amplitude function $\Xi(t)$ can be ignored within the time scale of the synchrotron oscillation period, analogous to the derivation with the linear approximation, we can derive the detuning property from Eq. (\ref{eq22_sec2}) as
\begin{equation}
    \sqrt{\left<\xi^2\right>}\approx \xi_0\sqrt{1+\left(\frac{\alpha\sp{\prime}}{\Delta\omega+\frac{1}{8}\Omega_s\omega^2\Xi^2(t)}\right)^2},
    \label{eq26_sec2}
\end{equation}
where $\alpha\sp{\prime}=\frac{\Omega_s \epsilon\sp{\prime}}{4}$. Eq. (\ref{eq26_sec2}) shows the negative detuning character, which depends on the oscillation amplitude. Negative detuning occurs when the beam is modulated and the oscillation amplitude does not diverge due to the negative detuning process. Although the detuning frequency is adjusted to diverge $\sqrt{\left< \xi^2 \right>}$ in Eq. (\ref{eq26_sec2}), namely $\Delta \omega = -\frac{1}{8}\Omega_s \omega^2 \Xi^2(t)$, such a condition leads to additional negative detuning due to an increase in $\Xi(t)$. Thus, the spontaneous detuning process results in a fixed bunch length that elongates without diverging.

Now, we approximate the oscillation amplitude $\left|\Xi(t)\right|$ that is given by the modulated bunch length $\sigma_\text{mod}$ under phase modulation as $\sqrt{2}\sigma_\text{mod}$. As the same as Eq. (\ref{eq15_sec2}), we write Eq. (\ref{eq26_sec2}) with the modulated and natural bunch length $\sigma_\text{mod}$ and $\sigma_\tau$, namely
\begin{equation}
    \sigma_\text{mod} \approx \sigma_\tau
    \sqrt{1+\Biggl(
    \frac{\alpha\sp{\prime}}{\Delta \omega+\frac{1}{4}\Omega_s \bigl(\omega \sigma_\text{mod}\bigr)^2}\Biggr)^2}.
    \label{eq27_sec2}
\end{equation}
As discussed in the linear approximation, the detuning frequencies of positive and negative sides $\Delta \omega_{\rho, \pm}$ where the modulated bunch length $\sigma_\text{mod}$ increases to $\rho$ times the natural bunch length $\sigma_\tau$ are given as
\begin{equation}
    \Delta\omega_{\rho,\pm} =
    \pm\frac{\alpha}{\kappa \sqrt{\rho^2-1}} - \frac{1}{4}\kappa \omega_s \left(\omega \rho \sigma_\tau \right) ^2, 
    \label{eq28_sec2}
\end{equation}
%
%

%\sout{Furthermore, we discuss the detuning frequency width $\Delta\omega_{\rho}$ where the averaged amplitude $\sqrt{\left<\xi^2\right>}$ is equal to $\rho\xi_0$ $(\rho>1)$. The positive and negative detuning frequencies $\Delta\omega_{\rho, \pm}$ where the bunch length increases to $\rho$ times the natural bunch length $\sigma_\tau$ are given as
%
%
% \begin{equation}
%     \frac{\Delta\omega_{\rho,\pm}}{\omega_s} =
%     \pm\frac{\phi_{m0}\cot \phi_0}{2\kappa \sqrt{\rho^2-1}} - \frac{1}{4}\kappa \left(\omega \sigma_m \right) ^2, 
%     \label{eq24_sec2}
% \end{equation}
% }
%
%
where $\kappa$ is the non-linear correction factor
\begin{equation}
\kappa = \sqrt{1-\frac{1}{4}\left(\omega \rho \sigma_\tau \right)^2}.
\label{eq29_sec2}
\end{equation}
%
%
%\sout{Especially, the detuning frequencies where the bunch length lengthens to $\sqrt{2}$ times the natural bunch length are written as 
%
%
% \begin{equation}
%     \frac{\Delta\omega_{\sqrt{2},\pm}}{\omega_s} =
%     \pm\frac{1}{2\kappa}\phi_{m0}\cot \phi_0- \frac{1}{4}\kappa \left(\omega \sigma_m \right) ^2. 
%     \label{eq26_sec2}
% \end{equation}
% %
% %
% Eqs. (\ref{eq24_sec2}) and (\ref{eq26_sec2}) show a negative detuning character, and the detuning frequency depends on the bunch length under phase modulation, being enhanced under larger modulation amplitude conditions. From Eq. (\ref{eq26_sec2}), the detuning frequency width where the bunch length lengthens to $\sqrt{2}$ times the natural bunch length is written as
% %
% %
% \begin{equation}
%   \frac{\Delta\omega_{\sqrt{2},+}}{\omega_s} - \frac{\Delta\omega_{\sqrt{2},-}}{\omega_s} = \frac{1}{\kappa}\phi_{m0} \cot \phi_0.
%   \label{eq27_sec2}
% \end{equation}
% %
% %
% }
With the non-linear approximation, the detuning frequency width $\Delta f_{\sqrt{2}}$, in which the modulated bunch length is larger than $\sqrt{2}\sigma_\tau$, is given as
\begin{equation}
    \Delta f_{\sqrt{2}} = \frac{\alpha}{\kappa_{\sqrt{2}} \pi},
    \label{eq30_sec2}
\end{equation}
where $\kappa_{\sqrt{2}} = \sqrt{1-\frac{1}{2}\left(\omega \sigma_\tau\right)^2}$. This shows that the detuning frequency width becomes large compared to the linear approximation. The second term on the right side of Eq. (\ref{eq28_sec2}) shows the characteristic of negative frequency detuning. The negative frequency shift $\delta f_{\sqrt{2}}$ in the case of $\rho=\sqrt{2}$ is given as
\begin{equation}
    \delta f_{\sqrt{2}} = -\frac{1}{4\pi} \kappa_{\sqrt{2}} \omega_s \left(\omega \sigma_\tau\right)^2.
    \label{eq31_sec2}
\end{equation}

To compare the linear and nonlinear approximations, we approximate the correction factor $\kappa$ as
\begin{equation}
\kappa^{\pm1} \approx 1\mp\frac{1}{8}\left(\omega \rho \sigma_\tau\right)^2.
\label{eq32_sec2}
\end{equation}
With this approximation, the detuning frequencies in Eq. (\ref{eq28_sec2}) are written as 
\begin{equation}
    \begin{aligned}
    \Delta\omega_{\rho,+} &=
    \ \frac{\alpha}{\sqrt{\rho^2-1}}
    -\frac{1}{4} \left( \omega \rho \sigma_\tau \right)^2
    \left(\omega_s-\frac{\alpha}{2\sqrt{\rho^2-1}} \right)
    \\
    &+\frac{1}{32}\left(\omega \rho \sigma_\tau\right)^4,
    \label{eq33_sec2}
    \end{aligned}
\end{equation}
\begin{equation}
    \begin{aligned}
    \Delta\omega_{\rho,-} &=
    \ -\frac{\alpha}{\sqrt{\rho^2-1}}
    -\frac{1}{4} \left( \omega \rho \sigma_\tau \right)^2
    \left(\omega_s+\frac{\alpha}{2\sqrt{\rho^2-1}} \right)
    \\
    &+\frac{1}{32}\left(\omega \rho \sigma_\tau\right)^4.
    \label{eq34_sec2}
    \end{aligned}
\end{equation}

Eqs. (\ref{eq33_sec2}) and (\ref{eq34_sec2}) show that the detuning curve from the linear approximation shifts in the negative frequency direction, and the frequency shift differs between the positive and negative sides. Accordingly, the detuning curve from the nonlinear approximation has a negative detuning character and an asymmetric form around the peak frequency.

As a measure, we estimate the frequency width of the detuning curve in the KARA case. Now, we assume that the total accelerating voltage $V_c$, the radiation loss per turn $U_0$, and the synchrotron angular frequency $\omega_s$ are 1.36 MV, 622.4 keV, and $2\pi\times$34.3 kHz, which are the values we have evaluated from the experiment mentioned in section \ref{Experiment}. Regarding the phase modulation amplitude, 10.25 degrees was one of the modulation amplitude values in the experiments. As mentioned in section \ref{Experiment}, we activated phase modulation at one of the two RF stations. In this case, the parameter $\phi_{m0}$ in the model must be half of the set experimental value. That is, $\phi_{m0}$ must correspond to $\frac{1}{2}\times 10.25=5.13$ degrees. From these parameters, we evaluate the frequency width $\Delta f_{\sqrt{2}} = 0.74$ kHz from Eq. (\ref{eq30_sec2}). Concerning the negative frequency shift due to nonlinearity, we evaluate the negative detuning frequency $\delta f_{\sqrt{2}} = -0.38$ kHz from Eq. (\ref{eq31_sec2}).

\begin{figure}
\includegraphics[width=\columnwidth]{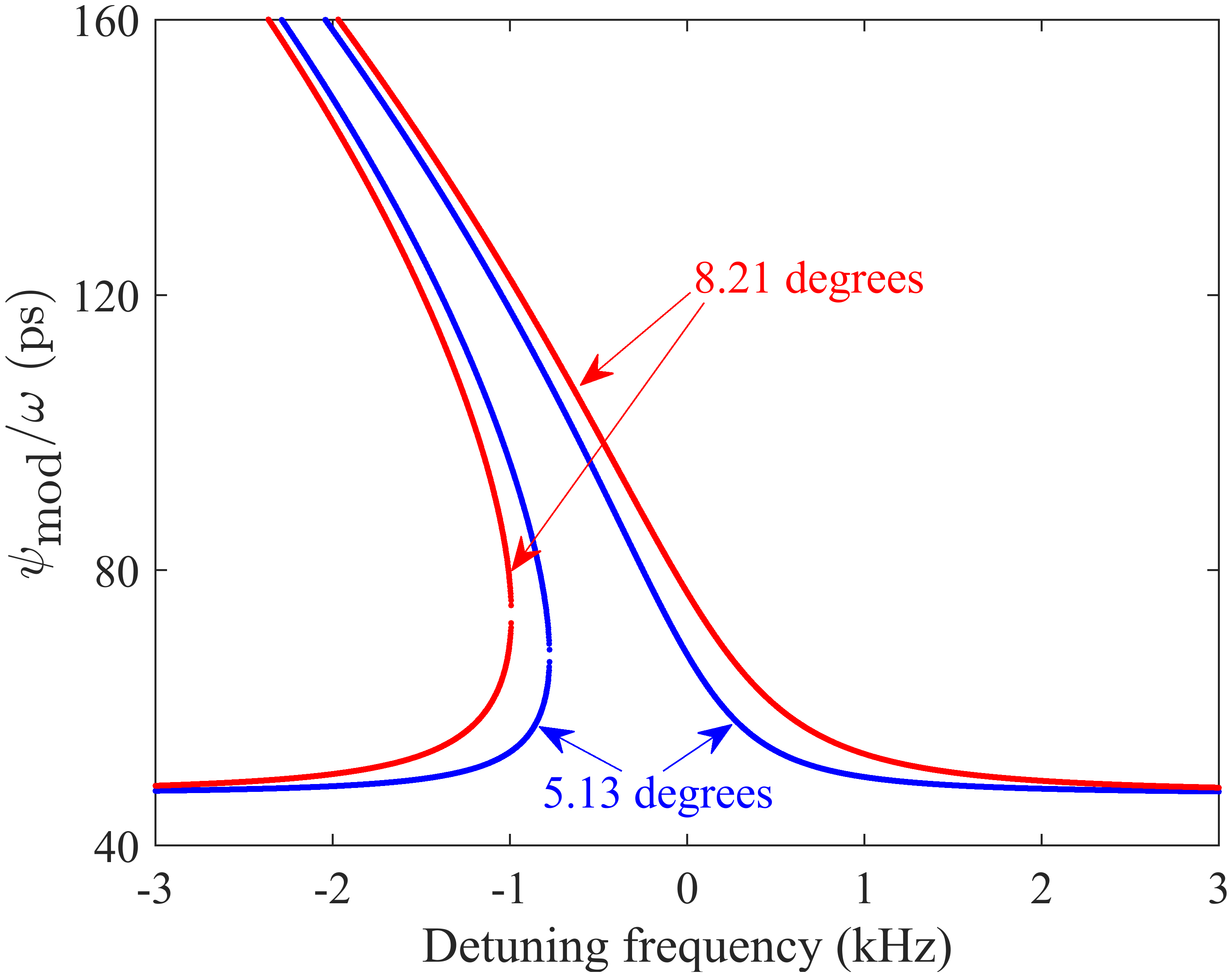}
\caption{\label{fig1}
The numerical results correspond to the physical solutions of Eq. (\ref{eq35_sec2}) for two modulation amplitudes: 5.13 and 8.21 degrees. These results correspond to the frequency-detuning characteristics of bunch lengthening due to phase modulation. 
}
\end{figure}

\begin{figure}
\includegraphics[width=\columnwidth]{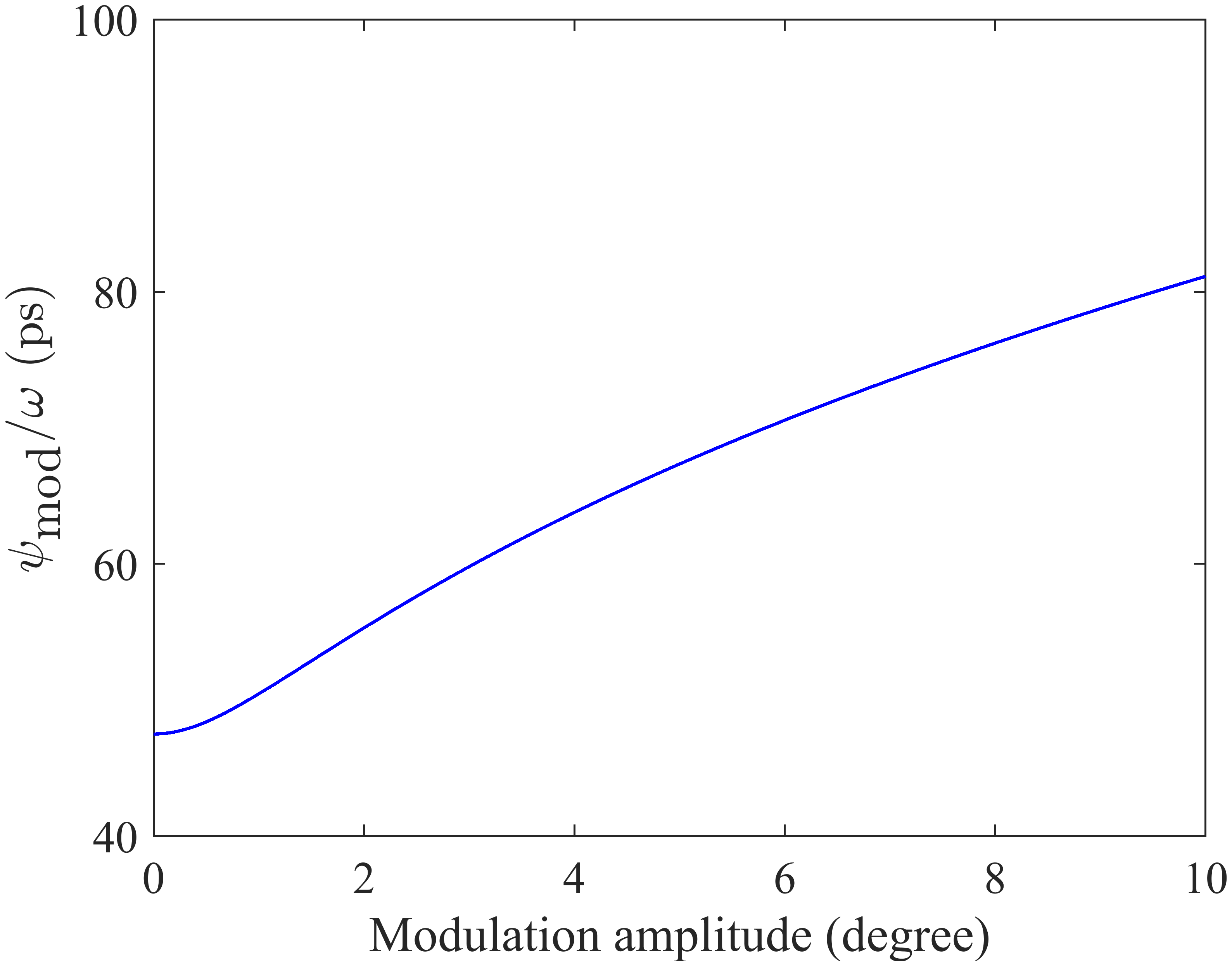}
\caption{\label{fig2}
The dependence of the positive solution of Eq. (\ref{eq36_sec2}) on the phase modulation amplitude. This corresponds to elongated bunch lengths, which can be achieved with the given phase modulation amplitudes.}
\end{figure}

The Equation (\ref{eq27_sec2}) also shows the relationship between the modulated bunch length and the frequency detuning, namely the detuning character. We write the phases occupied by the natural bunch length $\sigma_{\tau}$ and by the modulated bunch length $\sigma_{\text{mod}}$ as $\psi_{\tau} = \omega \sigma_{\tau}$ and $\psi_{\text{mod}} = \omega \sigma_{\text{mod}}$. 
For simplicity, assuming that the phase occupied by the modulated bunch length is small enough to approximate as $1-\left(\frac{\psi_{\text{mod}}}{2}\right)^2\approx 1$, Eq. (\ref{eq27_sec2}) can be written as the following cubic equation for $\psi^2_{\text{mod}}$: 
\begin{equation}
\begin{aligned}
    &\left(\psi^2_\text{mod}\right)^3 +\left(8\frac{\Delta \omega}{\omega_s}-\psi^2_\tau \right)\left(\psi^2_\text{mod}\right)^2 \\
    &+8\frac{\Delta \omega}{\omega_s}\left(2\frac{\Delta \omega}{\omega_s}-\psi^2_\tau\right) \psi^2_\text{mod} \\
    &-16\left( \left(\frac{\Delta \omega}{\omega_s}\right)^2+\left(\frac{\alpha}{\omega_s}\right)^2\right)\psi^2_\tau=0.
    \label{eq35_sec2}
    \end{aligned}
\end{equation}
For a value of $\sigma_{\text{mod}}=100$ ps, the value $1-\left(\frac{\psi_{\text{mod}}}{2}\right)^2$ is about 0.97, which shows the validity of the approximation used in Eq. (\ref{eq35_sec2}). The values of $\psi^2_{\text{mod}}$, which are the solutions to the cubic equation (\ref{eq35_sec2}) that have positive values, are related to the bunch length. 
Figure \ref{fig1} shows the numerical results of the modulated bunch length $\sigma_{\text{mod}}=\frac{\psi_{\text{mod}}}{\omega}$ for two different modulation amplitude cases, namely $\frac{1}{2}\times 10.25=5.13$ and $\frac{1}{2}\times 16.42 = 8.21$ degrees by assuming the same KARA operation parameters as in the discussion about the frequency width and the negative detuning frequency. As can be seen in the figure, the detuning curves separate into two parts, indicating the negative detuning character with an asymmetric form due to nonlinearity. The results also show that the bunch length does not diverge as the detuning frequency approaches zero due to the nonlinear nature of the detuning curve.  
%
%
%As seen in Eq. (\ref{eq27_sec2}), negative detuning occurs spontaneously under the nonlinear approximation. Due to this spontaneous detuning, the oscillation amplitude of the electron does not diverge even though the detuning frequency is zero. Therefore, we can apply $\Delta \omega = 0$ to Eq. (\ref{eq27_sec2}) and obtain the following equation
%
%
% \begin{equation}
%  \psi_\text{mod}^2 = \psi_\tau^2 \Biggl( 1+\frac{\epsilon^2}{\psi^4_\text{mod} \Bigl(1-\bigl(\frac{1}{2}\psi_\text{mod}\bigr)^2 \Bigr)^2} \Biggr),
%  \label{eq35_sec2}
% \end{equation}
%
%
%where $\psi_\text{mod}=\omega \sigma_\text{mod}$ and $\psi_\tau=\omega \sigma_\tau$. Since the phase $\psi_\text{mod}$ occupied by the bunch is small, we can approximate as $1-\left(\frac{1}{2}\psi_\text{mod}\right)^2\approx 1$. Consequently, Eq. (\ref{eq35_sec2}) can be written as the following cubic equation for $\psi_\text{mod}^2$,
As a special case, we obtain the following cubic equation for $\psi^2_{\text{mod}}$ in the case of $\Delta \omega =0$, whose solution gives the typical dependence of bunch lengthening on modulation amplitude: 
\begin{equation}
    \left(\psi_\text{mod}^2\right)^3-\psi_\tau^2 \left( \psi_\text{mod}^2\right)^2-\epsilon^2 \psi^2_\tau=0.
    \label{eq36_sec2}
\end{equation}
The solution to Eq. (\ref{eq36_sec2}) that has a positive value gives the bunch length that can be achieved with phase modulation. Figure \ref{fig2} shows the dependence of the length of the elongated bunch $\sigma_\text{mod}=\frac{\psi_\text{mod}}{\omega}$ on the phase modulation amplitude derived from Eq. (\ref{eq36_sec2}). The results show an elongated bunch length of 67.8 ps for a phase modulation amplitude of 5.13 degrees at $\Delta \omega = 0$, whereas the natural bunch length is 47.5 ps. However, as seen in Fig. \ref{fig1}, the bunch length can exceed the values shown in Fig. \ref{fig2} because the detuning curve significantly tilts towards the negative frequency direction. 
%
%

%%%%%%%%%%%%%%%%%%%%%%%%%%%%%%%%%%%%%%%%%%%%%%%%%%%%%%%%%%%%%%%%%%%%%%%%%%%%%%%%%%%%%   
\section{\label{Experiment}Experiment}
The experiment was carried out at the Karlsruhe Research Accelerator (KARA) electron storage ring. Table \ref{table1} summarizes the main machine parameters of KARA for the normal user operation mode at 2.5 GeV. In the experiment, we operated the KARA storage ring with the same bunch-filling pattern as the regular users' operations, which had serial three-bunch trains and a gap with empty RF buckets. The total number of filled RF buckets is typically 99, and the gap following the bunch trains has 85 empty RF buckets. 
\begin{table}[b]
\caption{\label{table1}
Main parameters of KARA storage ring.
}
\begin{ruledtabular}
\begin{tabular}{lccc}
\textrm{Parameter} & {Symbol} & {Value} & {Unit} \\
\colrule
Beam energy & $E_0$ & 0.5 - 2.5 & GeV \\
RF frequency & $f_\text{rf}$ & 499.7 & MHz \\
Harmonic number & $h$ & 184 & \\
Total accelerating voltage \\(maximum) & $V_c$ & 1.6 & MV \\
Momentum compaction factor \\ (designed at 2.5 GeV) & $\alpha$ & 0.00867 & \\
Radiation loss per turn & $U_0$ & 622.4 & keV \\
Natural energy spread & $\frac{\sigma_\delta}{E_0}$ & $9.08\times 10^{-4}$ & \\
%Natural bunch length & $\sigma_t$ & 36.9 & ps \\
%Beam current (typ.) & $I_b$ & 150 & mA \\
Beam current (typical) & $I_b$ & 150 & mA \\
Number of RF sections & & 2 & \\
Number of cavities per RF section & & 2 & \\
\end{tabular}
\end{ruledtabular}
\end{table}
To carry out the RF phase modulation at KARA, we have introduced the modulation function into our low-level RF (LLRF) system (Dimtel, LLRF9/500 \cite{DIMTELwebsite}). Each RF station in the KARA storage ring has one klystron, whose maximum output power is 250 kW. Because one RF station has two RF accelerating cavities, one klystron drives two RF cavities in KARA. Each RF station has its own LLRF system to control two RF stations independently. From one master oscillator, whose signal frequency is around 499.7 MHz, each LLRF system in each RF station receives the master clock signal and drives the RF station. Because the LLRF system has the function of phase modulation, we can perform phase modulation for each RF station independently. In the experiments discussed in the following part, we performed phase modulation at only one RF station, which is the same condition as in the simulation discussed in section \ref{Simulation}. To perform the phase modulation, we remotely set and measure the modulation frequency and amplitude using the LLRF system. The LLRF system has a function for measuring the modulation pattern of the forward power to the cavities. By analyzing the modulation pattern, we measured the modulation amplitude and frequency. The phase modulation amplitude values we applied in the experiments were $10.25\pm0.37$, $13.45\pm0.56$, and $16.42\pm1.36$ degrees, with the errors in the modulation phase arising from reproducibility. As discussed in section \ref{Simulation}, the simulation results agree with the experiments and show slightly smaller modulation phase amplitudes than the values obtained from the forward power measurement. This suggests the possibility that the modulation amplitude in the RF cavity is smaller than that in the forward power.
The technical details for the method of phase modulation by the LLRF system are discussed in \cite{TeytelmanARXIV2019}.

To observe the longitudinal motion of the beam, we use a visible light streak camera (VSC, Hamamatsu, C5680-21S) at the visible light diagnostic beamline (VLD) \cite{KehrerIPAC2015} at KARA. In the VLD, we can use the synchrotron radiation (SR) beam that comes from a 5-degree port of one of the bending magnets at KARA. In the front-end part of the beamline, we use one mirror chamber to filter out the high-energy component of the SR beam and to extract the beam from the ultra-low vacuum environment to the atmosphere. Therefore, we can use the visible light component of the SR beam at the VLD in the atmospheric environment. The streak camera measurement needs to achieve excellent temporal resolution, with a typical time scale of several to ten picoseconds. To do so, we use a focusing mirror system to avoid aberration and deliver the spot image into the input slit of the VSC. The VSC has two sweeping units: a fast sweeping unit (synchro scan unit) and a slow sweeping unit (blanking amplifier unit). By the fast sweeping unit, the VSC can observe the pulse profile of the SR beam, which corresponds to the electron bunch profile in the storage ring, with synchronization between the sweep timing of the VSC and the revolution timing of the electron beam. From this, the electron bunch length can be measured. By the slow sweeping unit, the VSC can also observe the temporal change in the longitudinal bunch profile, whose time scale is typically longer than the revolution period. Because these two sweeping units function simultaneously, we can precisely observe both the bunch length and the global longitudinal motion of the beam.
\subsection{\label{Experiment_ModulationFrequencyAmplitude}Modulation frequency and amplitude}
\begin{figure}
\begin{tabular}{l}
(a) \\
\includegraphics[width=\columnwidth]{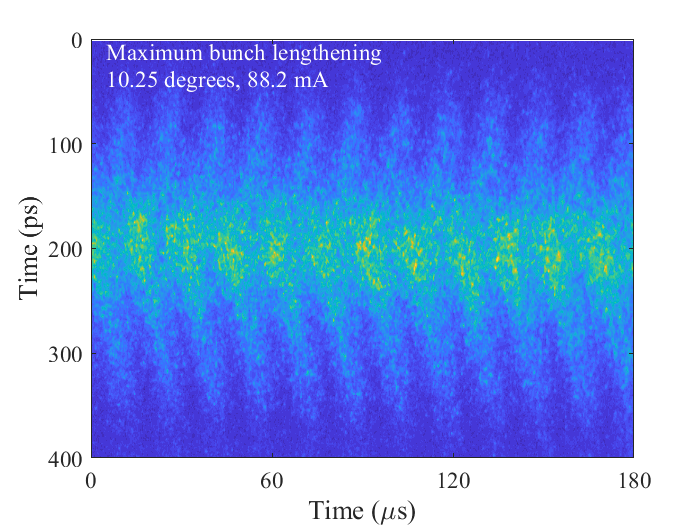}
\\
(b) \\
\includegraphics[width=\columnwidth]{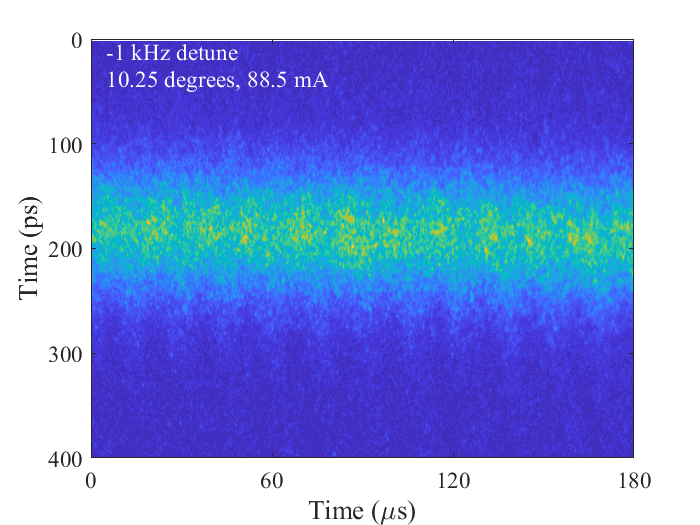}
\\
(c) \\
\includegraphics[width=\columnwidth]{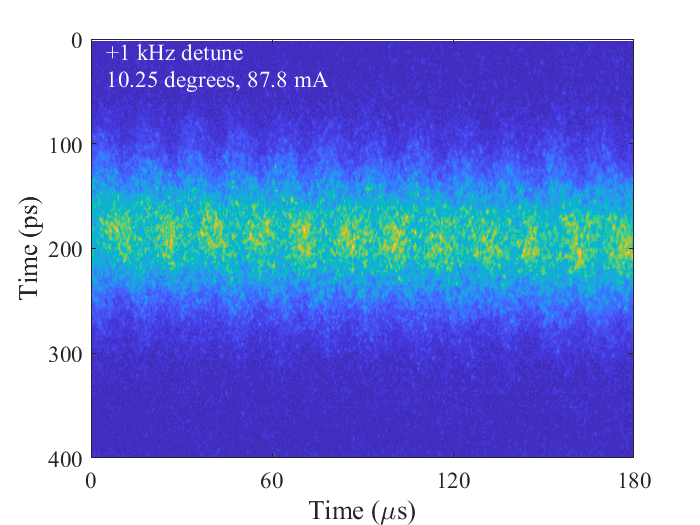}
\end{tabular}
\caption{\label{fig3} 
The experimental results of the streak camera measurement, which show the change in the longitudinal beam profile. (a) shows the result at the maximum bunch lengthening condition with a modulation frequency of 67.0 kHz. (b), (c) show the streak camera results at modulation frequencies of 66.0 and 68.0 kHz, which correspond to the frequency conditions of -1 and +1 kHz from the maximum bunch lengthening condition. The beam current and the modulation amplitude are 88 mA and 10.25 degrees.}
\end{figure}

We have measured the bunch lengthening while changing the modulation frequency and amplitude of the phase modulation. Figures \ref{fig3}(a)-\ref{fig3}(c) show the streak camera images measured with a modulation amplitude of 10.25 degrees and a beam current of 88 mA. Figure \ref{fig3}(a) shows the streak camera image for the modulation frequency that causes the maximum bunch lengthening. Figures \ref{fig3}(b) and \ref{fig3}(c) show the images with -1 kHz and +1 kHz detuning from the modulation frequency in Fig. \ref{fig3}(a). The modulation frequency change of $\pm$1 kHz from the maximum elongation condition in Fig. \ref{fig3}(a) clearly reduces the effect of bunch lengthening. As seen in these figures, the electron bunch length oscillates over time and the oscillation period is almost the same as half the synchrotron oscillation period $(\sim 15\ \mu \text{s})$.

\begin{figure}
\includegraphics[width=\columnwidth]{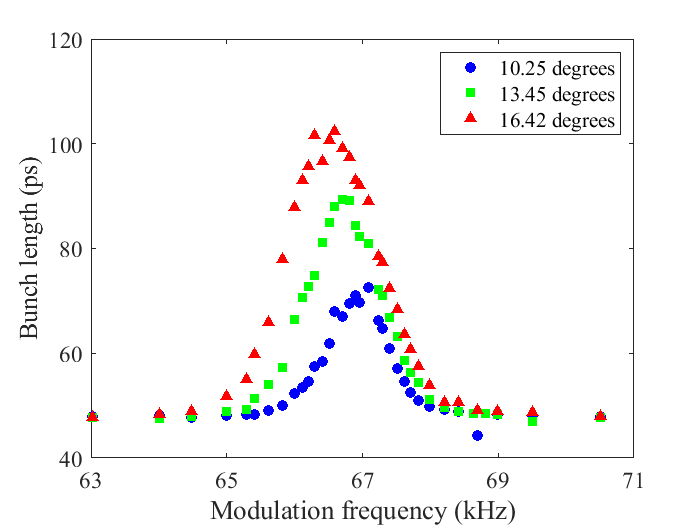}
\caption{\label{fig4}
The experimental result of the detuning curve for the bunch lengthening under phase modulation with different sets of modulation amplitudes: 10.25 (blue circles), 13.45 (green squares), and 16.42 degrees (red triangles). 
These measurements were conducted with a beam current of 90 mA. The horizontal axis corresponds to the modulation frequency, which is close to twice the synchrotron frequency \cite{MaierMaster}.}
\end{figure}

By changing the modulation frequency and amplitude systematically, we analyzed the detuning curve of the bunch length \cite{MaierMaster}. 
Figure \ref{fig4} shows the detuning curves of the bunch length with modulation amplitudes of (10.25, 13.45, 16.42) degrees and a beam current of 88 mA. 
To improve the statistics, we took 100 image data at each modulation frequency condition, and all of the data points in Fig. \ref{fig4} show the statistical errors; however, the statistical error values are smaller than the marks in this figure. 
From the detuning curves, we evaluated the peak frequency width and compared it with the theoretical value. 
The peak width values of each detuning curve, defined as the frequency region where the bunch length is larger than $\sqrt{2}\sigma_\tau$, are (0.55, 1.37, 1.89) kHz for (10.25, 13.45, 16.42) degrees modulation amplitudes. 
According to the theory, when considering that the phase modulation was activated at one of the two RF stations, the estimated peak width values for each modulation amplitude are (0.79, 1.04, 1.27) kHz. These values are comparable with the experimental results.
%\sout{As we mentioned in Sec. \ref{TheoreticalModel}, the derivation of the detuning curve does not consider the radiation damping effect. 
%This simplification leads to an overestimation of the theoretical values and the discrepancy between the theoretical estimations and the experimental results.} 

As shown in Fig. \ref{fig4}, the peak frequency shifts in the negative direction with larger modulation amplitudes. 
The peak frequency values are (66.93, 66.72, 66.55) kHz for the modulation amplitudes of (10.25, 13.45, 16.42) degrees, respectively. 
This negative shift aligns with the theoretical consideration in which negative frequency detuning, depending on the phase-modulated bunch length, is discussed.

\begin{figure}
\includegraphics[width=\columnwidth]{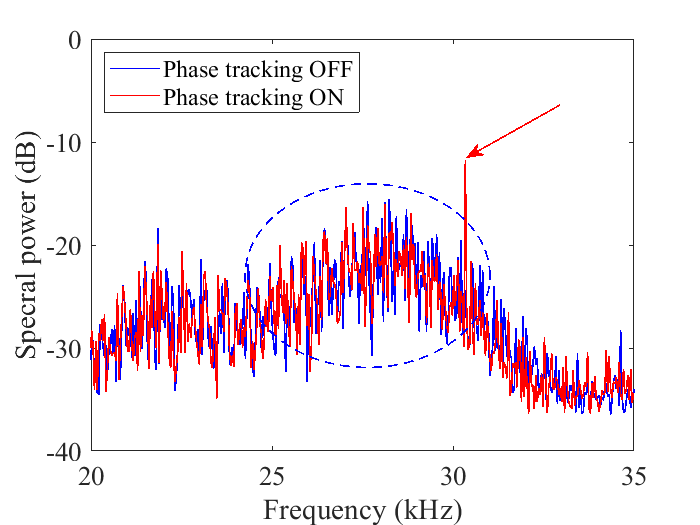}
\caption{\label{fig5}
The longitudinal beam spectrum with and without phase tracking (red and blue curves). 
The broad peak (indicated by the blue ellipse) appears independently of phase tracking. 
The sharp peak (indicated by the red arrow) on the red curve appears only when phase tracking is on and has a higher frequency than the broad peak.}
\end{figure}

To discuss the negative detuning character with the measured detuning curve in depth, we measured the synchrotron oscillation frequency $f_s$ in parallel with the bunch length measurement by measuring the longitudinal frequency spectrum using the bunch-by-bunch feedback system \cite{BlomleyIPAC2016}. 
The result of the measured synchrotron frequency was $(32.6\pm0.54)$ kHz, where the measurement error is due to the uncertainties in the frequency readout. 
This meant that the detuning curves in Fig. \ref{fig4}, even though in the most prominent negative detuning case with 16.42 degrees of modulation, did not show the negative detuning character discussed in section \ref{TheoreticalModel}. 
However, we found that the readout value of the bunch-by-bunch feedback system had a systematic error by comparing the measured synchrotron frequency with a phase tracking method \cite{ZhangIBIC2020, BlomleyPhDThesis}, which allows us to measure the synchrotron frequency precisely. 
Therefore, we conducted another measurement to evaluate the systematic difference in the synchrotron frequency between with and without phase tracking. 
Figure \ref{fig5} shows a typical longitudinal spectrum measured with the bunch-by-bunch feedback system with and without phase tracking. 
The beam spectrum has a broad peak, which we analyzed as the synchrotron oscillation spectrum and is indicated by the blue ellipse in the figure, and a sharp peak indicated by the arrow in the figure that always appears in a higher frequency region than the broad peak when phase tracking is on. 
We analyzed the difference in frequency between the broad and sharp peaks and treated it as a systematic error in the measurement of the synchrotron frequency. 
We repeated the measurement with several different RF accelerating voltages and beam current conditions while keeping a multi-bunch and 2.5 GeV condition. 
From the measurement, the systematic error of the measured synchrotron frequency was evaluated to be $(+1.73 \pm 0.36)$ kHz. 
From the estimation, the double synchrotron frequency, including the systematic error, was evaluated as $(68.6\pm1.8)$ kHz. 
The detuning curve of 16.42 degrees in Fig. \ref{fig4} peaks around 66.6 kHz, indicating negative detuning from the double synchrotron oscillation period, considering the error. 
As discussed in section \ref{TheoreticalModel}, the negative detuning frequencies, assuming the phase-modulated bunch length lengthens to $\sqrt{2}$ times the natural bunch length and the maximum bunch length with a modulation amplitude of 16.42 degrees in Fig. \ref{fig4}, are estimated to be around -0.4 and -0.9 kHz, which are smaller than the readout error value of the synchrotron frequency in this experiment. 
The largest modulation phase amplitude case in this experiment indicates only the observable negative frequency detuning.

According to the theoretical model, the achievable bunch length with phase modulation can be evaluated by solving Eq. (\ref{eq36_sec2}). For modulation amplitudes of $\frac{1}{2}\times$(10.25, 13.45, 16.42) degrees, the solutions to Eq. (\ref{eq36_sec2}) are the bunch lengths of (67.8, 72.7, 76.8) ps, respectively, whereas the peak values in the measured frequency detuning curves in Fig. \ref{fig4} are (72.6, 89.4, 102.3) ps for the same modulation amplitudes. As discussed in section \ref{TheoreticalModel}, the solution to Eq. (\ref{eq36_sec2}) tends to underestimate the bunch length due to the tilt of the detuning curve in Fig. \ref{fig1}. In this case, the underestimation is approximately 30 \%.
\subsection{\label{Experiment_BeamCurrent}Beam current dependence}
\begin{figure}
\includegraphics[width=\columnwidth]{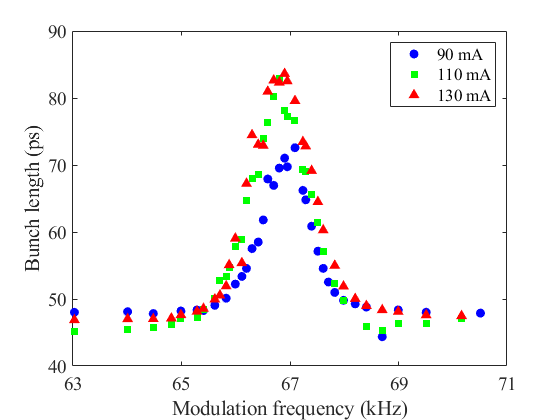}
\caption{\label{fig6}
The experimental results of the detuning curve for the bunch lengthening under phase modulation with the same modulation amplitude (10.25 degrees) and different beam current conditions: 90 mA (blue circles), 110 mA (green squares), and 130 mA (red triangles). }
\end{figure}

To study the beam loading effect on bunch lengthening with phase modulation, we observed the bunch length and the detuning curves while changing the beam current. 
Figure \ref{fig6} shows three detuning curves with different beam currents while keeping the modulation amplitude at the same value (10.25 degrees). 
As shown in the figure, bunch lengthening due to phase modulation tends to be slightly more effective under higher beam current conditions. 
According to the theoretical model in Eqs. (\ref{eq7_sec2}) and (\ref{eq22_sec2}), the factors $\epsilon$ and $\epsilon'$, which show the magnitude of phase modulation, depend on the synchronous phase and increase when the RF accelerating voltage decreases. 
The LLRF system regulates the klystron power to stabilize the RF accelerating voltage by compensating for the beam loading effect. 
However, due to the partial filling pattern at KARA, a transient beam loading effect that occurs during the passage of bunch trains cannot be perfectly compensated. 
A slight change in the accelerating voltage occurs under different beam current conditions due to the transient beam loading effect, which causes the dependence of the phase modulation effect on the beam current. 

Another possible reason for the current dependence of the bunch length is a potential well distortion 
\cite{book_Chao} due to short-range wakefields. In the KARA case, the bunch lengthening due to potential well distortion is negligible in the bunch current range shown in Fig. \ref{fig6}. The difference in the maximum elongated bunch length between 90 and 110 mA is shown as around 10 ps. In the bunch length in the detuning frequency regions where the modulation frequency is far from the peak frequency or the double synchrotron frequency, there is less systematic dependence on the bunch length with the beam current.  This indicates that there is less systematic difference in the less modulated bunch length within the beam current range shown in Fig. \ref{fig6}.
%\sout{In the KARA case, however, the bunch lengthening due to the potential well distortion is estimated to be negligible with the 2.5 GeV multi-bunch operation, assuming the resistive wall impedance. 
%The estimation of the bunch lengthening due to potential well distortion with the resistive wall impedance is summarized in Appendix \ref{appendix3}. %}
%
%

\subsection{\label{Experiment_BeamLifetime}Beam lifetime}
\begin{figure}
\includegraphics[width=\columnwidth]{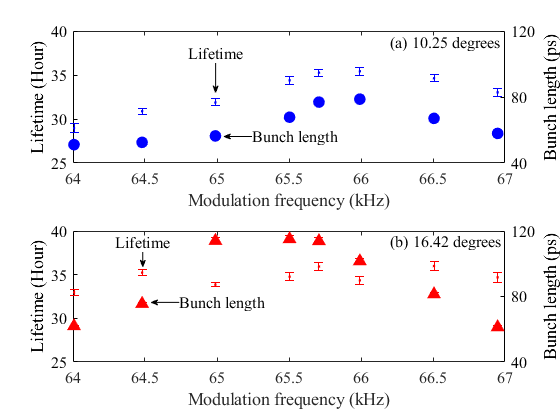}
\caption{\label{fig7}
The measured beam lifetime and bunch length with the modulation amplitudes of (a) 10.25 and (b) 16.42 degrees at the same beam current of 107 mA.}
\end{figure}

One of the benefits of RF phase modulation is an improvement in beam lifetime. 
By changing the modulation frequency, we observed another detuning curve regarding the measured beam lifetime. 
Figure \ref{fig7} shows the detuning curve for the beam lifetime and the bunch length with modulation amplitudes of 10.25 and 16.42 degrees under the same beam current condition of 107 mA. 
For the 10.25-degree modulation amplitude, the bunch length maximized around the 66 kHz modulation frequency, and the beam lifetime reached approximately 35 hours at that frequency. This increase in beam lifetime is 16 \% higher than the beam lifetime without phase modulation (30 hours).
%\sout{In the case of the 10.25-degree modulation amplitude, the beam lifetime reached around 35 hours at the peak of the detuning curve, which corresponds to a 16\% increase in the beam lifetime compared to the lifetime without modulation (30 hours).} 

On the other hand, the bunch length showed maximum lengthening at the 65 kHz modulation frequency for the 16.42-degree modulation amplitude. However, the beam lifetime did not increase beyond the 10.25-degree case. Additionally, the beam lifetime showed less dependency on the modulation frequency and less correlation with the bunch lengthening.
%\sout{On the other hand, in the case of the 16.42-degree modulation amplitude, the beam lifetime reached the same value as in the 10.25-degree case. 
% However, the beam lifetime did not increase beyond the 10.25-degree case and showed less dependency on the modulation frequency.}  
This implies that the Touschek scattering effect does not determine the beam lifetime due to the elongation of the bunch length but is determined by other beam loss processes, such as the beam loss due to scattering processes between the electron beam and residual gas molecules in the vacuum chambers of the accelerator.

The slight difference in modulation frequency is visible at the maximum bunch length condition when comparing the bunch lengthening between the 110 mA case in Fig. \ref{fig6} and Fig. \ref{fig7}(a). The bunch length reaches its maximum value at 66.8 kHz in Fig. \ref{fig6} and around 66 kHz in Fig. \ref{fig7}(a).
It should be noted that the measurements in Fig. \ref{fig7} took place as another experiment, separately from the measurements in Fig. \ref{fig6}.

However, as shown in Fig. \ref{fig7}, we confirmed experimentally that the beam lifetime was improved by RF phase modulation. 

%This implies an over-excitation in the 16.42-degree case, by which the electron beam is lost through some dynamic processes, such as the horizontal dynamic aperture effect. 
%The phase modulation modulates both the bunch length and the energy spread, and the transverse beam size can change due to the non-zero dispersion function at the KARA storage ring. 
%The discussion about transverse beam dynamics is outside the scope of this paper. However, we confirmed that there is an optimized modulation amplitude to improve beam lifetime.  
%
%
%
%

%%%%%%%%%%%%%%%%%%%%%%%%%%%%%%%%%%%%%%%%%%%%%%%%%%%%%%%%%%%%%%%%%%%%%
\section{\label{Simulation}Simulation}
We have performed numerical simulations based on macro-particle tracking to investigate beam dynamics under phase modulation with frequency detuning. 
The method used is based on \cite{SakanakaPRSTAB2000}. 
We have calculated the equations of motion for the longitudinal direction stated in Eqs. (\ref{eq1_sec2}) and (\ref{eq2_sec2}) numerically while considering the beam loading effect, which arises from the beam-induced voltage in each RF accelerating cavity in the KARA storage ring, taking into account the filling time for each cavity. 
In the simulation, we have treated the RF accelerating voltage $V_c=\left| \vec{V_c} \right|$ as the vector sum of the generator voltage $\vec{V_g}$ and the beam-induced voltage $\vec{V_b}$, namely, $\vec{V_c}=\vec{V_g} + \vec{V_b}$. 
$\vec{V_g}$ has a direct effect on phase modulation, and both $\vec{V_c}$ and $\vec{V_b}$ have changed due to the phase modulation accordingly. 
To calculate the generator voltage $\vec{V_g}$ and the beam-induced voltage $\vec{V_b}$, we have used the measured values of the shunt impedance, the unloaded quality factor, and the input coupling value for each cavity, which were measured at their installation into the KARA storage ring.
In our simulation, we have used the definition of the shunt impedance $R_s$ as follows:
\begin{equation}
    R_s = \frac{V_c^2}{P_c},
    \label{eq1_sec4}
\end{equation}
where $P_c$ is the input power to the cavity. 
In the simulation, we have used individual shunt impedance values for each cavity: 6.64, 7.04, 7.00, and 6.04 M$\Omega$, respectively.

To consider the beam loading effect, we have treated both the phase advance and the amplitude of the beam-induced voltage $\vec{V_b}$ between adjacent bunches in multi-bunch filling conditions using the same method as in \cite{SakanakaPRSTAB2000}. 
Because of the characteristics of the beam loading effect, initially, $\vec{V_b}$ inside each cavity can change with the beam revolution step. 
Therefore, we have performed enough revolution steps in the simulation to achieve a quasi-stable condition in which $\vec{V_b}$ and $\vec{V_c}$ can oscillate slightly but do not change drastically. 
Typically, we have performed the revolution step over a period of up to three times the radiation damping time. 
To get a realistic solution for the motion of the electron, we have introduced the radiation excitation into Eqs. (\ref{eq1_sec2}) and (\ref{eq2_sec2}) as a random walk of the energy spread, which has a Gaussian distribution with a standard deviation $\sigma_r$ as \cite{SakanakaPRSTAB2000}
\begin{equation}
\sigma^2_r = 4\gamma_\epsilon T_0 \left( \frac{\sigma_\delta}{E_0}\right)^2,
\label{eq2_sec4}
\end{equation}
where $\sigma_\delta$ is the natural energy spread and is $9.08\times 10^{-4}$ in the KARA 2.5 GeV case.

We evaluated the relevant parameters from the experimental data to conduct the simulation, which we can directly compare to the experimental results. 
Especially, the values of the momentum compaction factor and the RF accelerating voltage significantly impact the simulation results because these parameters primarily affect the longitudinal beam dynamics.
The momentum compaction factor was evaluated using the measured synchrotron oscillation frequency and the bunch length mentioned in section \ref{Experiment}. 
The momentum compaction factor $\alpha_c$ is written as 
\begin{equation}
    \alpha_c = \sigma_\tau \omega_s\left(\frac{\sigma_\delta}{E_0}\right)^{-1}.
    \label{eq3_sec4}
\end{equation}
We evaluated the momentum compaction value with the measured synchrotron oscillation frequency and the measured bunch length when the phase modulation was off. 
Considering the uncertainty of the synchrotron oscillation frequency, the momentum compaction factor was evaluated as $\alpha_c = 1.126\times 10^{-2} \pm 2.95\times 10^{-4}$.

Concerning the RF accelerating voltage, it is possible that the value differs from the scalar sum of the two accelerating voltage values at the two RF stations due to the phases at each RF station. 
The total RF accelerating voltage seen by the beam, $V_c$, is written as
\begin{equation}
    V_c = \frac{1}{e}\sqrt{
    \left(\frac{2\pi h E_0 \omega_s}{\sigma_\tau \omega_\text{rf}^2}\right)^2
    \left(\frac{\sigma_\delta}{E_0}\right)^2
    + U_0^2},
    \label{eq4_sec4}
\end{equation}
and the error $\Delta V_c$ arising from the uncertainty in the measured synchrotron oscillation frequency $\Delta \omega_s$ is
\begin{equation}
    \Delta V_c = \frac{(eV_c)^2-U_0^2}{e^2V_c}\frac{\Delta \omega_s}{\omega_s}.
    \label{eq5_sec4}
\end{equation}
From these two equations, the total RF accelerating voltage seen by the beam was evaluated as $V_c=1.362 \pm 0.028$ MV. 
In the following discussion, we used these values as the momentum compaction factor and the RF accelerating voltage to reproduce the experiment in the simulation. 
To compare the simulation results with the experiment, we applied the same bunch-filling pattern in the simulation as in the experiment mentioned in section \ref{Experiment}.

We conducted the simulation with 10000 macro-particles to observe snapshots of the longitudinal phase space distribution. 
We also conducted the tracking simulation by scanning phase modulation parameters such as modulation frequency and amplitude. 
In such scanning simulations, we reduced the number of macro-particles to 2,000 to save calculation time. 
We confirmed that the results don't significantly change with different numbers of macro-particle settings.

In the simulation, we considered three tuning parameters of RF phase modulation: modulation frequency, modulation amplitude, and beam current. 
Regarding the modulation amplitude, we found that the simulation results enhance the lengthening effect under the same phase modulation amplitudes measured by the cavity's forward power in the experiments. 
In the simulations, we applied (7.91, 11.78, 14.80) degrees as the phase modulation amplitudes, instead of (10.25, 13.45, 16.42) degrees in the experiments. This adjustment of the amplitude values is made to ensure consistency between the simulation and the experiment. In the following, we discuss the dependence of bunch lengthening on these parameters.

\subsection{\label{Simulation_ModulationFrequency}Modulation frequency and amplitude}
\begin{figure}
\includegraphics[width=\columnwidth]{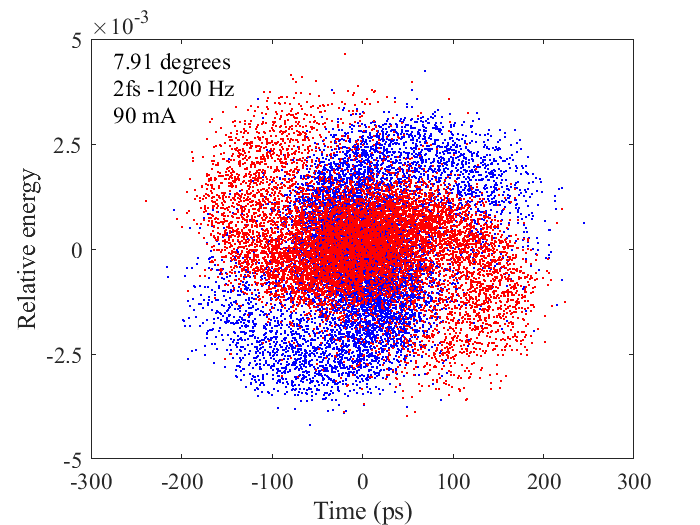}
\caption{\label{fig8}
The simulation result of the longitudinal phase space distribution when the RF phase modulation is active with a total beam current of 90 mA. 
The modulation frequency and amplitude are set to $2 f_s-1200$ Hz and 7.91 degrees, respectively. 
The blue and red dots correspond to snapshots of the electron distribution with a time difference of a quarter of one synchrotron oscillation period.}
\end{figure}

Figure \ref{fig8} shows the simulation results of longitudinal phase space snapshots in the case of active RF phase modulation, where the modulation amplitude is 7.91 degrees and the modulation frequency is adjusted to maximize the bunch length condition of $2f_s-1200$ Hz. 
The total beam current is set at 90 mA in the simulation. 
The time difference between these two snapshots in Fig. \ref{fig8} is a quarter of one synchrotron oscillation period. 
These results show that both the bunch length and the energy spread change with phase modulation. 
Two spiral arms emerge from the beam core, indicating nonlinear effects.

\begin{figure}
\begin{tabular}{l}
(a) \\
\includegraphics[width=\columnwidth]{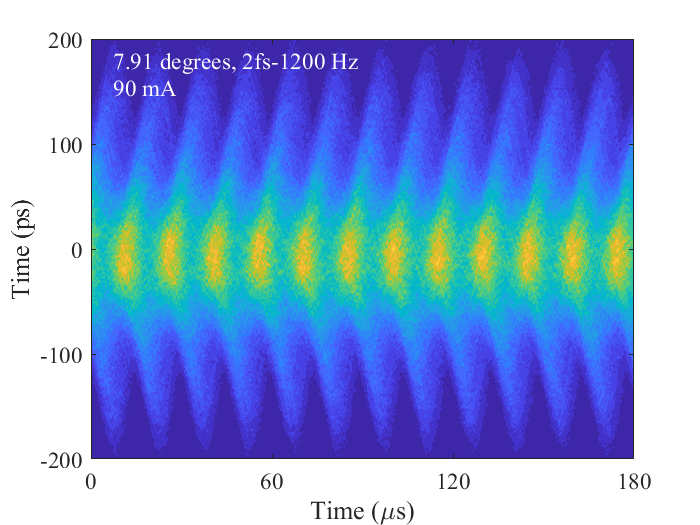}
\\
(b) \\
\includegraphics[width=\columnwidth]{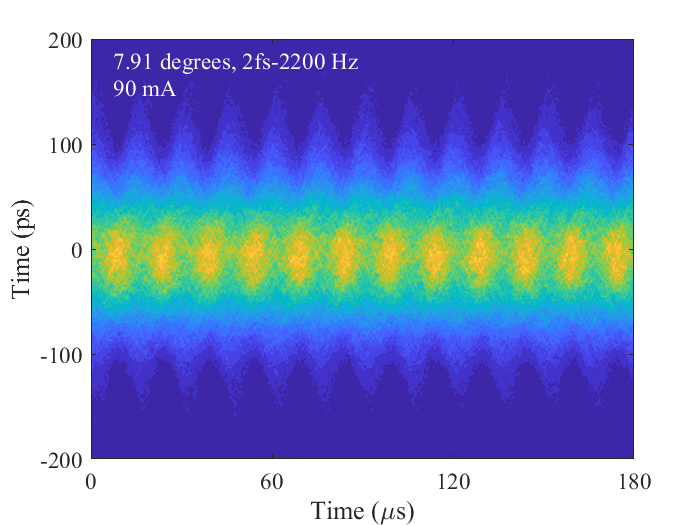}
\\
(c) \\
\includegraphics[width=\columnwidth]{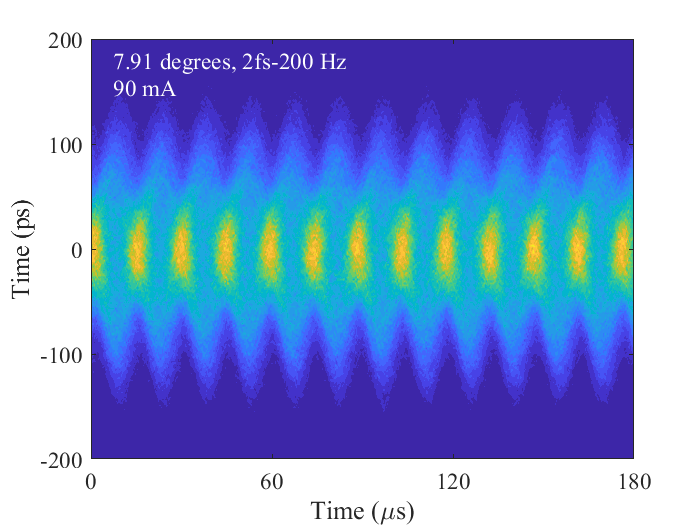}
\end{tabular}
\caption{\label{fig9}
The simulation results of the change in the longitudinal electron distribution in one bunch when the modulation frequency is set to maximize the bunch lengthening [$2 f_s -1200$ Hz;(a)], -1 kHz, and +1 kHz detuning from condition (a) [(b), (c)]. 
The 90 mA total beam current is set in the simulation, and all the other machine parameters are the same as in Fig. \ref{fig8}. The vertical scale measures the bunch length. The positive direction on the vertical axis corresponds to the head of the bunch. The blue and yellow colors indicate low and high electron density in the charge distribution, respectively.}
\end{figure}

Figure \ref{fig9} (a) is a color map showing the change in the longitudinal electron distribution in one bunch for a typical time scale of the synchrotron oscillation period when the modulation frequency is set to maximize the bunch lengthening shown in Fig. \ref{fig8}. 
We can see that the simulated electron beam behaves similarly to the experimental result in Fig. \ref{fig3}(a), where the modulation frequency is adjusted to maximize the bunch lengthening.  
Figs. \ref{fig9} (b) and \ref{fig9}(c) show the color maps when the modulation frequency is shifted to 1 kHz in both negative and positive directions from the maximum bunch length condition. 
These simulation results also indicate a similarity to the experimental results shown in Figs. \ref{fig3}(b) and \ref{fig3}(c).

\begin{figure}
\includegraphics[width=\columnwidth]{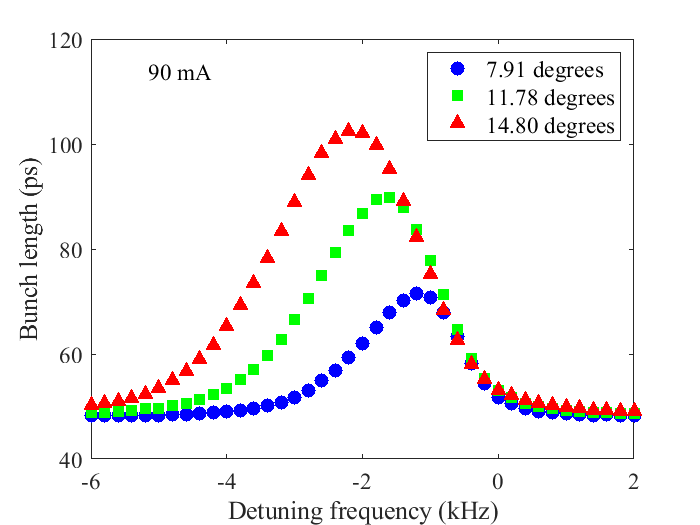}
\caption{\label{fig10}
The simulation results of the detuning curve for the modulation amplitude of 7.91 (blue circles), 11.78 (green squares), and 14.80 degrees (red triangles) at a total beam current of 90 mA. 
The horizontal axis corresponds to the frequency difference between $2 f_s$ and the modulation frequency.}
\end{figure}

By scanning the modulation frequency, we have summarized the dependence of bunch lengthening on the detuning frequency under different modulation amplitude conditions. 
Figure \ref{fig10} shows three frequency-detuning curves of bunch lengthening with modulation amplitudes of (7.91, 11.78, 14.80) degrees. 
To analyze the bunch length in Fig. \ref{fig10}, we have evaluated the time-averaged bunch length because it oscillates with time, as seen in Figs. \ref{fig9} (a)-(c).

We evaluated the frequency width of the detuning curve, where the bunch length is larger than $\sqrt{2}\sigma_\tau$. 
The simulation results in Fig. \ref{fig10} show the frequency widths of (0.78, 2.20, 3.01) kHz for the modulation amplitudes of (7.91, 11.78, 14.80) degrees, which are comparable to the experimental results: (0.55, 1.37, 1.89) kHz with modulation amplitudes of (10.25, 13.45, 16.42) degrees. 
On the other hand, the frequency width from the nonlinear approximation in Eq. (\ref{eq30_sec2}) is estimated to be 
%\sout{$|2\Delta \omega_{\sqrt{2}}|/2\pi=$}
$\Delta f_{\sqrt{2}} =$ (0.61, 0.91, 1.15) kHz for $\frac{1}{2}\times$(7.91, 11.78, 14.80) degrees, which are smaller than the simulation results. The difference between the simulation and the theoretical model tends to be more pronounced in cases with larger amplitudes. This suggests that the approximations treated in the theoretical model cause the discrepancy. Regarding the negative frequency detuning, the peak detuning frequencies for each modulation amplitude condition in Fig. \ref{fig10} are (-1.2, -1.6, -2.2) kHz for (7.91, 11.78, 14.80) degrees, respectively. The theoretical model, on the other hand, estimates the detuning frequency as (-0.43, -0.67, -0.87) kHz for the same three amplitude conditions by using the maximum bunch lengths for each modulation amplitude condition in Fig. \ref{fig10} to evaluate the negative detuning frequency. The theoretical model also estimates negative detuning frequencies to be smaller than the simulation results.
%(1.23, 1.83, 2.29) kHz for (7.91, 11.78, 14.80) degrees, which is greater than the simulation results.
% %\sout{ 
% The discrepancy, as discussed in Sec. \ref{Experiment_ModulationFrequencyAmplitude}, originates from a simplification in the theoretical model, where the radiation damping effect is not considered in the derivation of the detuning curve. }

The maximum bunch length is (71.45, 89.73, 102.34) ps for the modulation amplitudes of (7.91, 11.78, 14.80) degrees in Fig. \ref{fig10}, while the theoretical model from Eq. (\ref{eq36_sec2}) evaluates the achievable bunch length as (63.6, 70.21, 74.6) ps for the same modulation amplitude settings. As discussed in section \ref{TheoreticalModel}, these results underestimate the simulation results; however, the difference between the theoretical estimation and the simulation results is approximately 30 \%.
%According to the theoretical model, the negative detuning frequency $\delta f_{\text{mod}}$ with the modulated bunch length $\sigma_{\text{mod}}$ is written as
%
%
% \begin{equation}
% \begin{aligned}
%     \delta f_{\text{mod}} &= -\frac{1}{8\pi}\kappa_{\text{mod}} \omega_s \left(\omega \sigma_{\text{mod}}\right)^2, 
%     \label{eq6_sec4}\\
%     \kappa_{\text{mod}} &= \sqrt{1-\frac{1}{4}\left(\omega \sigma_{\text{mod}}\right)^2}.
% \end{aligned}
% \end{equation}
%
%
%From Eq. (\ref{eq6_sec4}), the detuning frequencies for the modulation amplitudes of (7.91/2, 11.78/2, 14.80/2) degrees are evaluated to be (-0.34, -0.41, -0.47) kHz. These values are smaller than the negative frequency detuning in Fig. \ref{fig10} for the same modulation amplitudes: (-1.2, -1.6, -2.2) kHz. }}
%\sout{Figure \ref{fig9} shows the peak frequency detuning in the negative direction. 
% Namely, the peak frequency shifts from $2f_s$ for each modulation amplitude are (-1.2, -1.6, -2.2) kHz at (7.91, 11.78, 14.80) degrees. 
% The peak frequency shifts by 0.4 kHz between 7.91 and 11.78 degrees, and by 1.0 kHz between 7.91 and 14.80 degrees. 
% These frequency changes at different modulation amplitudes are comparable to the experimental results in Fig. \ref{fig3}. 
% The theoretical model estimates that the negative detuning frequencies are (-0.43, -0.67, -0.87) kHz for modulation amplitudes of (7.91, 11.78, 14.80) degrees. 
% This underestimation indicates prominent nonlinear effects in the simulation that are not properly treated by the theoretical model. } 
%
%
\begin{figure}
\includegraphics[width=\columnwidth]{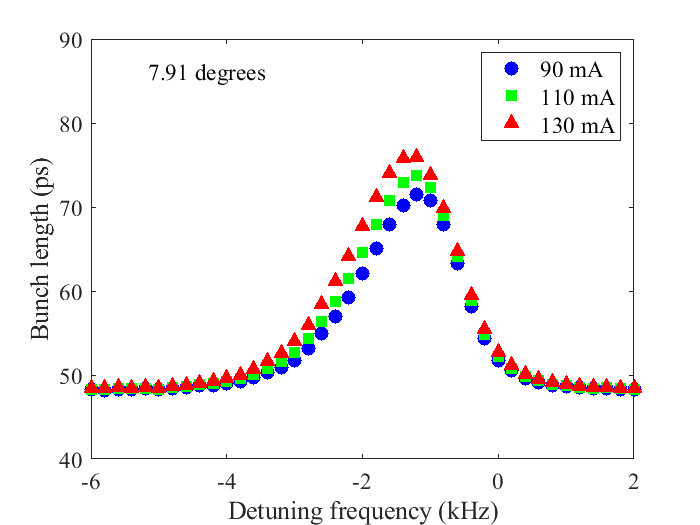}
\caption{\label{fig11}
The simulation results of the detuning curve for three different total beam current conditions with the same modulation amplitude of 7.91 degrees. 
Blue circles, green squares, and red triangles correspond to 90, 110, and 130 mA conditions, respectively.}
\end{figure}

\subsection{\label{Simulation_BeamCurrent}Beam current}
In the simulation, we have considered the beam loading effect that occurs inside the RF cavities. 
Therefore, the bunch lengthening depends on the beam current. 
Figure \ref{fig11} shows the detuning curves for three different beam current conditions, each with the same modulation amplitude of 7.91 degrees. 
As shown in the figure, the bunch lengthening increases under high-beam current conditions due to phase modulation. 
This beam current dependence agrees with the experimental results in Fig. \ref{fig6}. 
According to Eq. (\ref{eq7_sec2}), the driving term from the phase modulation is proportional to the parameter $\epsilon$, which includes $\cot \phi_0$. 
As discussed in section \ref{Experiment_BeamCurrent}, a transient beam loading effect occurs when bunch trains pass through, causing a slight change in the accelerating voltage.
%Due to beam loading, \sout{the synchronous phase $\phi_0$ changes because of energy loss.} 
This change reduces the over-voltage factor, which leads to an increase in the phase modulation amplitude parameter, $\epsilon$ in Eq. (\ref{eq7_sec2}). 
However, the beam current dependence is not significant, despite the difference in beam current, because the bunch currents for each multi-bunch filling condition are not widely different.

\section{\label{Summary}Summary}
In high-energy electron storage rings dedicated to high-energy physics and synchrotron radiation experiments, it is essential to prepare stable beams for the experiments. 
One way to stabilize the beam and improve its quality is to enhance the beam lifetime by lengthening the electron bunch. 
In this paper, we discussed the beam dynamics of RF phase modulation, by which the bunch length can change considerably. 
One of the critical parameters for lengthening the bunch efficiently is the modulation frequency. 
The modulation frequency should be close to twice the synchrotron oscillation frequency, $2f_s$. 
To discuss the detuning properties of bunch lengthening, we have considered a theoretical model that treats the frequency difference between the modulation frequency and $2f_s$.
The theoretical model shows that, with linear approximation, the oscillation amplitude of electrons diverges when the detuning frequency is zero. However, with a nonlinear approximation, the oscillation amplitude does not diverge, even when the detuning frequency is zero, because the resonance frequency spontaneously shifts in a negative direction due to the nonlinear effect. Consequently, the bunch length elongates but does not diverge resonantly under phase modulation.
The analytical form of the detuning curve shows that the width of the detuning curve depends on the modulation amplitude and the synchronous phase. The peak frequency of the detuning curve shifts in the negative frequency direction with the asymmetric peak form due to nonlinear effects. 
With the theoretical model, we discussed the elongated bunch length with phase modulation and derived an equation whose solution provides the achievable bunch length with phase modulation.

To consider these theoretical results and discussions experimentally, we have introduced the phase modulation function into our low-level RF system and conducted experiments at the KARA storage ring with 2.5 GeV electron beams. 
In the data analyses, we have introduced the detuning curve to discuss the dependence of bunch lengthening on the modulation frequency, which can be directly compared with the theoretical model by analyzing the width of the detuning curve. 
The experimental results showed that the frequency width widens at greater phase modulation amplitudes, which corresponds to the theoretical model. 
%However, the theoretical model overestimates the frequency width compared to the experimental results. 
%\sout{This indicates that the radiation damping effect, which is not accounted for in deriving the detuning curve in the theoretical model, makes the frequency width narrower.}  
The theoretical model also shows the negative detuning character due to nonlinearity, which was not prominently observed in the experiments because of the experimental uncertainty in the synchrotron frequency. 
However, despite the uncertainty, we have verified the negative detuning characteristic in the experiments.
In the experiment, we observed that the bunch lengthening effect increased with a larger phase modulation amplitude. Theoretical evaluations showed the same tendency and agreed with the experimental results, with approximately 30 \% despite the simplification.

To consider the beam dynamics of phase modulation in depth, we performed a simulation based on the macro-particle tracking method. 
By considering the equation of motion for the longitudinal motion of electrons, including the effects of RF phase modulation and beam loading inside the RF accelerating cavities, we calculated the behavior of the electron distribution in longitudinal phase space. 
From the results, we derived the detuning property of bunch lengthening using the detuning curves and verified the agreement of the detuning curve width with the experimental results. 
%However, the theoretical model overestimates the frequency width under the simulated conditions, reflecting the same tendency observed in the comparison with the experiments. 
The simulation results showed a negative detuning characteristic, which corresponded to the experimental results. 
The results of bunch elongation from the simulation agreed with the theoretical evaluations with simplification.
%The theoretical model underestimated negative frequency detuning under the simulated condition, indicating strong nonlinear effects in the simulation, which are not treated properly by the theoretical model.
The experiments and the simulation results showed that the beam loading effect occurring in the RF accelerating cavity impacts bunch lengthening due to phase modulation. 
However, the effect was not remarkable in the case of the KARA storage ring.

One of the benefits of RF phase modulation is an improvement in beam lifetime because the Touschek lifetime depends on the electron density in a bunch. 
We observed the beam lifetime with two different phase modulation amplitudes under the same beam current conditions. 
At the 10.25-degree condition, the beam lifetime peaked while scanning the modulation frequency where the bunch length also had the peak value. 
%\sout{exhibiting the same tendency as the detuning curve of bunch lengthening.} 
At the 16.42-degree condition, the beam lifetime also improved up to the value that the 10.25-degree condition reached and did not show any peak characteristics, although the detuning curve of the bunch length showed a peak. This implies that the Touschek lifetime improved due to bunch elongation with RF phase modulation, and the beam lifetime is determined by other beam loss processes, such as scattering between the electron beam and the residual gas molecules, and there are fewer benefits to increasing the modulation amplitude more than 16.42 degrees in this case. 

\appendix
\section{\label{appendix1}Derivation of detuning character}
To solve Eq. (\ref{eq13_sec2}), we assume the following forms as the solutions $a_1(t)$ and $b_1(t)$ by introducing two functions $C(t)$ and $D(t)$
\begin{eqnarray}
    a_1(t) &=& C(t) e^{-\frac{\alpha}{\Delta\omega}\cos\Delta\omega t},
    \label{eq1_app1} \\
    b_1(t) &=& D(t) e^{\frac{\alpha}{\Delta\omega}\cos\Delta\omega t}.
    \label{eq2_app1}
\end{eqnarray}
Because of Eq. (\ref{eq13_sec2}), $C(t)$ and $D(t)$ are the solutions of the following equations
\begin{eqnarray}
    \dot{C}(t) &=& -\alpha\cos\Delta\omega t \  e^{\frac{2\alpha}{\Delta\omega}\cos\Delta\omega t}D(t), 
    \label{eq3_app1} \\
    \dot{D}(t) &=& -\alpha\cos\Delta\omega t \ e^{-\frac{2\alpha}{\Delta\omega}\cos\Delta\omega t}C(t).
    \label{eq4_app1}
\end{eqnarray}
In a special case where $\alpha=0$, the functions $C(t)$ and $D(t)$ are constants. 
This means that $a_1(t)$ and $b_1(t)$ are also constants; accordingly, they are the solutions without phase modulation. 
In the following, we assume that $\alpha \neq 0$.

From Eq. (\ref{eq3_app1}) and (\ref{eq4_app1}), we get the following equations
\begin{equation}
\begin{aligned}
    \ddot{C}(t)+\bigl(\Delta\omega\tan\Delta\omega t&+2\alpha\sin\Delta\omega t\bigr) \dot{C}(t)\\
    &=\alpha^2\cos^2\Delta\omega t \ C(t), 
    \label{eq5_app1}
\end{aligned}
\end{equation}
\begin{equation}
\begin{aligned}
    \ddot{D}(t)+\bigl(\Delta\omega\tan\Delta\omega t&-2\alpha\sin\Delta\omega t\bigr) \dot{D}(t)\\
    &=\alpha^2\cos^2\Delta\omega t \ D(t).
    \label{eq6_app1}
    \end{aligned}
\end{equation}
The right-side term in Eqs. (\ref{eq5_app1}) and (\ref{eq6_app1}) proportional to $\alpha^2$ can be ignored because we assume a small RF phase modulation amplitude. 
Solving for $\dot{C}(t)$ and $\dot{D}(t)$ we get
\begin{eqnarray}
    \dot{C}(t) &=& C_0 \cos\Delta\omega t \ e^{\frac{2\alpha}{\Delta\omega}\cos\Delta\omega t}, 
    \label{eq7_app1} \\
    \dot{D}(t) &=& D_0 \cos\Delta\omega t \ e^{-\frac{2\alpha}{\Delta\omega}\cos\Delta\omega t},
    \label{eq8_app1}
\end{eqnarray}
where $C_0$ and $D_0$ are constants.

There is no solution in closed form for the open integral $\int \cos \theta e^{\pm \cos \theta}d\theta$. 
To obtain an analytical form, we expand the exponential part around zero and derive the $(\frac{2\alpha}{\Delta \omega}\cos \Delta \omega t)^k$ polynomial. 
%\sout{This means that we consider the behavior of the system, which is not close to the resonance condition where $\frac{2\alpha}{\Delta \omega}$ has a larger value.}
This means that we consider the behavior of the system where the value of $\frac{2\alpha}{\Delta \omega}$ is less than one. This situation occurs when the detuning frequency $\Delta \omega$ is larger than $2\alpha$, i.e., the far resonance condition. The situation also occurs when $\Delta \omega$ has a small value with small $\alpha$, i.e., the near resonance condition.
Taking up to the first-order term, we obtain
\begin{eqnarray}
    \dot{C}(t) &\approx& C_0\left(\cos\Delta\omega t + \frac{2\alpha}{\Delta\omega}\cos^2\Delta\omega t \right), \label{eq9_app1} \\
    \dot{D}(t) &\approx& D_0\left(\cos\Delta\omega t - \frac{2\alpha}{\Delta\omega}\cos^2\Delta\omega t \right). \label{eq10_app1}
\end{eqnarray}
The solutions for $C(t)$ and $D(t)$ are
\begin{equation}
\begin{aligned}
    C(t)&=C_0 
    \Biggl(
    \frac{\sin\Delta\omega t}{\Delta\omega}+\frac{\alpha}{\Delta\omega}\left(\frac{\sin 2\Delta\omega t}{2\Delta\omega}+t\right) 
    \Biggr)\\
    &+C_1, \label{eq11_app1}
\end{aligned}
\end{equation}
\begin{equation}
\begin{aligned}
    D(t)&=D_0
    \Biggl(\frac{\sin\Delta\omega t}{\Delta\omega}-\frac{\alpha}{\Delta\omega}\left(\frac{\sin 2\Delta\omega t}{2\Delta\omega}+t\right)\Biggr)\\
    &+D_1, \label{eq12_app1}
\end{aligned}
\end{equation}
where $C_1$ and $D_1$ are constants. 
To avoid the solution $a_1(t)=b_1(t)=0$ at the complete detuning condition with $\Delta\omega \rightarrow \pm \infty$, $C_1$ and $D_1$ should not be zero.

First, we discuss the behavior of the system under the far resonance condition.
With the solutions of $C(t)$ and $D(t)$, we can express the solutions of $a_1(t)$ and $b_1(t)$ from Eq. (\ref{eq1_app1}) and (\ref{eq2_app1}) using the expansion of the exponential part $e^{\pm \frac{\alpha}{\Delta\omega}\cos\Delta\omega t}\approx1\pm\frac{\alpha}{\Delta\omega}\cos\Delta\omega t$ and with dropping $\alpha^2$ terms 
\begin{equation}
\begin{aligned}
    a_1(t)&=\frac{1}{\Delta\omega}\left(C_0\sin\Delta\omega t-C_1\alpha\cos\Delta\omega t\right)\\
    &+C_0\frac{\alpha}{\Delta\omega}t+C_1, \label{eq13_app1}
\end{aligned}
\end{equation}
\begin{equation}
\begin{aligned}
    b_1(t)&=\frac{1}{\Delta\omega}\left(D_0\sin\Delta\omega t+D_1\alpha\cos\Delta\omega t\right)\\
    &-D_0\frac{\alpha}{\Delta\omega}t+D_1. \label{eq14_app1}
\end{aligned}
\end{equation}
To consider the frequency-detuning character, we introduce the oscillation amplitude $\xi(t)$ without the radiation-damping term
\begin{equation}
    \xi^2(t) = a_1^2(t)+b_1^2(t).
    \label{eq15_app1}
\end{equation}
The motion of the electron in Eq. (\ref{eq8_sec2}) is written with $\xi(t)$ as
\begin{equation}
    \tau = e^{-\gamma_\epsilon t} \xi(t) \sin\left(\omega_s t + \theta\right),
    \label{eq16_app1}
\end{equation}
where $\cos\theta = \frac{a_1(t)}{\sqrt{a_1^2(t)+b_1^2(t)}}$. 
The phase modulation affects the amplitude function $\xi(t)$, and the radiation damping term is separated from $\xi(t)$ in Eq. (\ref{eq16_app1}).

We consider the averaged $\xi^2(t)$ over the time period $\frac{2\pi}{\Delta\omega}$ 
\begin{equation}
\begin{aligned}
    \left<\xi^2\right>&\equiv\frac{\Delta\omega}{2\pi}\int^{\frac{2\pi}{\Delta\omega}+t_0}_{t_0}\xi^2(t) dt\\
    &=\frac{\Delta\omega}{2\pi}\int^{\frac{2\pi}{\Delta\omega}+t_0}_{t_0} \left(a_1^2(t)+b_1^2(t)\right)dt.
    \label{eq17_app1}
\end{aligned}
\end{equation}
where $t_0$ is the time when we observe the beam's motion. 
From these integrals, we get
\begin{equation}
\begin{aligned}
    \left<a_1^2\right>&=C_1^2+\frac{1}{2\Delta \omega^2}\left(C_0^2+\alpha^2 C_1^2\right)\\
    &+\frac{4\pi^2}{3}\frac{\alpha^2}{\Delta\omega^4}C_0^2\left(1+3t_0\frac{\Delta\omega}{2\pi}+3t_0^2\left(\frac{\Delta\omega}{2\pi}\right)^2\right)\\
    &+2\alpha C_0 C_1 \left(\frac{1}{\Delta\omega^2}+\Delta\omega t_0\right)\\
    &-\frac{2\alpha}{\Delta\omega^3}\left(C_0^2\cos\Delta\omega t_0 + C_0 C_1 \sin \Delta \omega t_0 \right),
    \label{eq18_app1}
\end{aligned}
\end{equation}
\begin{equation}
\begin{aligned}
    \left<b_1^2\right>&=D_1^2+\frac{1}{2\Delta \omega^2}\left(D_0^2+\alpha^2 D_1^2\right)\\
    &+\frac{4\pi^2}{3}\frac{\alpha^2}{\Delta\omega^4}D_0^2\left(1+3t_0\frac{\Delta\omega}{2\pi}+3t_0^2\left(\frac{\Delta\omega}{2\pi}\right)^2\right)\\
    &-2\alpha D_0 D_1 \left(\frac{1}{\Delta\omega^2}+\Delta\omega t_0\right)\\
    &+\frac{2\alpha}{\Delta\omega^3}\left(D_0^2\cos\Delta\omega t_0 + D_0 D_1 \sin \Delta \omega t_0 \right).
    \label{eq19_app1}
    \end{aligned}
\end{equation}
From the equation of motion (\ref{eq13_sec2}) with $t=\frac{2\pi n}{\Delta \omega}$ where $n$ is an integer, gives us $C_0=-\alpha D_1$ and $D_0=-\alpha C_1$. 
From these relations, we get $C_0 C_1 = D_0 D_1$. Therefore, $\left<\xi^2\right>$ is given as
\begin{equation}
\begin{aligned}
&\left<\xi^2\right>=\left<a_1^2\right>+\left<a_1^2\right> \\
&=C_1^2 + D_1^2
+\frac{1}{2\Delta \omega^2}\left(C_0^2+D_0^2+\alpha^2\left(C_1^2+D_1^2\right) \right) \\
&-\frac{2\alpha}{\Delta \omega^3}\left(C_0^2-D_0^2\right)\cos \Delta \omega t_0\\
&+\frac{4\pi^2\alpha^2}{3\Delta \omega^4}\left( \left(C_0^2+D_0^2\right)\left(1+3\frac{\Delta \omega t_0}{2\pi}+3\left(\frac{\Delta \omega t_0}{2\pi}\right)^2\right) \right).
\label{eq20_app1}
\end{aligned}
\end{equation}
In the case where the modulation frequency is completely detuned, namely $\Delta\omega \rightarrow \pm \infty$, the oscillation amplitude must correspond to the equilibrium status where the phase modulation doesn't affect the beam. 
This requires
\begin{equation}
    \lim_{\Delta\omega \to \pm \infty} \left<\xi^2\right> = \xi_0^2,
    \label{eq21_app1}
\end{equation}
%
%
%\textcolor{red}{\sout{where $\xi_0$ is a constant and is related to the natural bunch length.}}
where $\xi_0$ is the paramater related to the natural bunch length.
From this condition, we get $C_1^2+D_1^2=\xi_0^2$. 
Because of the symmetry of $a(t)$ and $b(t)$, $C_1^2$ and $D_1^2$ should be identical and are given as $C_1^2=D_1^2=\frac{1}{2}\xi_0^2$. 
From the values of $C_1^2$ and $D_1^2$, $C_0^2$ and $D_0^2$ are given as $C_0^2=D_0^2=\frac{1}{2}\alpha^2 \xi_0^2$. 
Therefore, $\left<\xi^2\right>$ is written as 
\begin{equation}
\begin{aligned}
\left<\xi^2\right>&=\xi_0^2\Biggl(1+\left(\frac{\alpha}{\Delta \omega}\right)^2\\
&+\frac{4\pi^2}{3}\left(\frac{\alpha}{\Delta \omega}\right)^4\biggl(1+3\frac{\Delta \omega t_0}{2\pi}+3\left(\frac{\Delta \omega t_0}{2\pi}\right)^2 \biggr) \Biggr).
\label{eq22_app1}
\end{aligned}
\end{equation}
The terms in Eq. (\ref{eq22_app1}), which depend on $t_0$, represent amplitude growth over time and affect the oscillation amplitude. 
However, these terms can be neglected in the current approximation because of the factor with the order of $\alpha^4$. 
With this approximation, the averaged square amplitude $\left<\xi^2\right>$ is expressed as
\begin{equation}
\left<\xi^2\right> \approx \xi_0^2\left(1+\left(\frac{\alpha}{\Delta \omega}\right) ^2\right),
\label{eq23_app1}
\end{equation}
which corresponds to Eq. (\ref{eq14_sec2}).

Next, we discuss the behavior of the system under the near resonence condition. The equations (\ref{eq11_app1}) and (\ref{eq12_app1}) can be rewritten as
\begin{equation}
\begin{aligned}
    C(t)&=C_0 t
    \Biggl(
    \frac{\sin\Delta\omega t}{\Delta\omega t}+\frac{\alpha}{\Delta\omega}\left(\frac{\sin 2\Delta\omega t}{2\Delta\omega t}+1\right) 
    \Biggr)\\
    &+C_1, \label{eq24_app1}
\end{aligned}
\end{equation}
\begin{equation}
\begin{aligned}
    D(t)&=D_0 t
    \Biggl(\frac{\sin\Delta\omega t}{\Delta\omega t}-\frac{\alpha}{\Delta\omega}\left(\frac{\sin 2\Delta\omega t}{2\Delta\omega t}+1\right)\Biggr)\\
    &+D_1. \label{eq25_app1}
\end{aligned}
\end{equation}
When $\Delta \omega t$ goes close to zero, $C(t)$ and $D(t)$ are written as
\begin{equation}
        C(t)= C_0\left(1+\frac{2\alpha}{\Delta \omega}\right)t + C_1, \label{eq26_app1}
\end{equation}
\begin{equation}
        D(t)= D_0\left(1-\frac{2\alpha}{\Delta \omega}\right)t + D_1. \label{eq27_app1}
\end{equation}
With Eq. (\ref{eq26_app1}) and (\ref{eq27_app1}), the value $\xi^2(t) = a^2_1(t)+b^2_1(t)$ is given as
\begin{equation}
        \xi^2(t) = \frac{\xi_0^2}{2} 
        \Bigl(F_+(t) e^{-\frac{2\alpha}{\Delta \omega}}
        +F_-(t) e^{\frac{2\alpha}{\Delta \omega}}
        \Bigr),
        \label{eq28_app1}
\end{equation}
where
\begin{equation}
    \begin{aligned}
    F_{+}(t) &= \biggl(1\pm\alpha \Bigl(1 + \frac{2\alpha}{\Delta \omega}\Bigr)t \biggr)^2, \\
    F_{-}(t) &= \biggl(1\pm\alpha \Bigl(1 - \frac{2\alpha}{\Delta \omega}\Bigr)t \biggr)^2,
    \label{eq29_app1}
    \end{aligned}
\end{equation}
which correspond to Eq. (\ref{eq18_sec2}) and (\ref{eq19_sec2}).

Next, we clarify the relationship between the parameter $\xi_0$ and the natural bunch length. Equations (\ref{eq1_sec2}) and (\ref{eq2_sec2}), which are the equations of motion of the electron for the longitudinal direction, do not include the term for radiation excitation. Consequently, the solution to these equations (\ref{eq16_app1}) is a harmonic oscillation whose amplitude damps exponentially with the radiation damping time.

If we consider radiation excitation, we can write the solution of electron's motion in the form of the longitudinal invariant of motion:
\begin{equation}
    \tau(t) = \alpha_c \sqrt{W_s} \sin \left(\omega_s t + \theta \right),
    \label{eq30_app1}
\end{equation}
where $W_s$ is the invariant that takes longitudinal radiation damping and excitation into account and is written as
\begin{equation}
    \begin{aligned}
        W_s &= \frac{2C_\gamma \gamma^2}{J_\epsilon \omega_s^2}\frac{\oint \frac{ds}{\rho^3(s)}}{\oint\frac{ds}{\rho^2(s)}}, \\
        C_\gamma &= \frac{55}{32\sqrt{3}}\frac{\hbar}{mc}, 
        \label{eq31_app1}
    \end{aligned}
\end{equation}
where $\gamma$ is the Lorentz factor of the electron, $J_\epsilon$ is the longitudinal damping partition number, $\hbar$ is the reduced Pranck constant, $m$ is the rest mass of the electron, $c$ is the speed of light, and $\rho(s)$ is the curvature of the beam orbit at the position $s$ in the storage ring. \\
The natural bunch length $\sigma_\tau$ is given by
\begin{equation}
    \sigma^2_\tau = \frac{\omega_s}{2\pi}\int_0^{\frac{2\pi}{\omega_s}}\tau^2(t) dt=\frac{1}{2}\alpha^2_c W_s.
    \label{eq32_app1}
\end{equation}
Consequently, $\tau(t)$ in Eq. (\ref{eq30_app1}) is written as
\begin{equation}
    \tau(t) = \sqrt{2}\sigma_\tau \sin \left(\omega_s t + \theta \right).
    \label{eq33_app1}
\end{equation}
Comparison between Eq. (\ref{eq16_app1}) and (\ref{eq33_app1}) gives the relation between the parameter $(\xi(t), \xi_0)$ and the bunch length $(\sigma_{\text{mod}}, \sigma_{\tau})$. From Eq. (\ref{eq23_app1}) we obtain 
\begin{equation}
    \sigma_{\text{mod}} \approx \sigma_{\tau} \sqrt{1+\left(\frac{\alpha}{\Delta \omega}\right)^2 },
    \label{eq34_app1}
\end{equation}
which corresponds to Eq. (\ref{eq15_sec2}).

\section{\label{appendix2}Nonlinear approximation}
In case the electron's amplitude is large, we expand the cosine part in Eq. (\ref{eq2_sec2}) for $\phi_m$ and $\omega\tau$ until the third-order terms. Namely,
\begin{eqnarray}
\cos(\phi_0&+&\phi_{m}-\omega\tau) \approx 
\left( 1-\frac{1}{2}\phi_m^2 \right) \cos\phi_0
\nonumber \\
&+&\left( \frac{1}{6}\phi_m^3 - \phi_m \right) \sin\phi_0 \nonumber \\
&+&\left( 
\sin\phi_0+\phi_m\cos\phi_0-\frac{1}{2}\phi_m^2\sin\phi_0
\right) \omega \tau \nonumber \\
&+&\frac{1}{2}\left( \phi_m \sin\phi_0-\cos\phi_0 \right) 
(\omega \tau)^2\nonumber \\
&-&\frac{1}{6}\sin\phi_0 (\omega \tau)^3.
\label{eq1_app2}
\end{eqnarray}
By assuming $\phi_m$ is small, we ignore the $\phi_m^2$ and $\phi_m^3$ terms in Eq. (\ref{eq1_app2}). 
Then the cosine in Eq. (\ref{eq2_sec2}) is given as
\begin{eqnarray}
\cos (\phi_0&+&\phi_{m}-\omega\tau) \approx 
\cos\phi_0 - \phi_m \sin\phi_0\nonumber \\ 
&+&\left( \sin\phi_0+\phi_m\cos\phi_0 \right) \omega \tau \nonumber \\
&+&\frac{1}{2}\left( \phi_m\sin\phi_0 - \cos\phi_0 \right) (\omega \tau)^2\nonumber \\
&-&\frac{1}{6}\sin\phi_0 (\omega \tau)^3. 
\label{eq2_app2}
\end{eqnarray}
The equation of motion with this cosine is written as 
\begin{eqnarray}
    \frac{d^2\tau}{dt^2}&+&2\gamma_\epsilon\frac{d\tau}{dt}
    +\omega_s^2\left(1+\phi_m\cos\phi_0\right)\tau
    \nonumber \\
    &+&\frac{1}{2}\omega_s^2\left(\omega\phi_m-\omega\cot\phi_0\right)\tau^2
    -\frac{1}{6}\omega^2\omega_s^2\tau^3
    \nonumber \\
    &=& \frac{\omega_s^2}{\omega}\phi_m.
    \label{eq3_app2}
\end{eqnarray}
For the same reason as in Eq. (\ref{eq6_sec2}), we can drop the right side of Eq. (\ref{eq3_app2}). 
The equation of motion is then
\begin{eqnarray}
    \frac{d^2\tau}{dt^2}&+&2\gamma_\epsilon\frac{d\tau}{dt}
    +\omega_s^2\left(1+\phi_m\cos\phi_0\right)\tau
    \nonumber \\
    &+&\frac{1}{2}\omega_s^2\omega\left(\phi_m-\cot\phi_0\right)\tau^2
    -\frac{1}{6}\omega_s^2\omega^2\tau^3
    \nonumber \\
    &=& 0.
    \label{eq4_app2}
\end{eqnarray}
Now we assume the solution given in Eq. (\ref{eq8_sec2}) and apply it to the $\tau^2$ and $\tau^3$ terms in Eq. (\ref{eq4_app2}). 
The result is
\begin{widetext}
\begin{eqnarray}
    \frac{d^2\tau}{dt^2}&+&2\gamma_\epsilon\frac{d\tau}{dt}
    +\omega_s^2\left(1-\frac{1}{8}\omega^2\Xi^2(t)+\phi_m\cot\phi_0 \right)\tau \nonumber \\
    &=&-\frac{1}{2}\omega_s^2\omega
    \left(\phi_m-\cot\phi_0\right)
    \left[\frac{1}{2}\Xi^2(t)+a(t)b(t)\sin 2\omega_s t-\frac{1}{2}(a^2(t)-b^2(t))\cos2\omega_st \right]
    \nonumber \\
    &-&\frac{1}{24}\omega^2\omega_s^2
    \left[a(t)(a^2(t)-3b^2(t))\sin3\omega_st+b(t)(3a^2(t)-b^2(t))\cos3\omega_s t \right],
    \label{eq5_app2}
\end{eqnarray}
\end{widetext}
where $\Xi(t)=\sqrt{a^2(t)+b^2(t)}$ is the oscillation amplitude of the electron.
There are no terms on the right side in Eq. (\ref{eq5_app2}) that include $\cos\omega_s t$ and $\sin\omega_s t$. 
Therefore, we can ignore these terms because they have less effects on the motion of the electron. 
The equation of motion can be written as
\begin{equation}
    \frac{d^2\tau}{dt^2}+2\gamma_\epsilon\frac{d\tau}{dt}
    +\omega_s^2\left(1-\frac{1}{8}\omega^2\Xi^2(t)+\phi_m\cot\phi_0 \right)\tau =0.
    \label{eq6_app2}
\end{equation}
By introducing the following values
\begin{equation}
    \Omega_s^2=\omega_s^2 \left(1-\frac{1}{8}\omega^2\Xi^2(t) \right), 
    \label{eq7_app2}
\end{equation}
the equation of motion is written as
\begin{equation}
    \frac{d^2\tau}{dt^2}+2\gamma_\epsilon\frac{d\tau}{dt}
    +\Omega_s^2\left(1 +\frac{\phi_m\cot\phi_0}{1-\frac{1}{8}\omega^2\Xi^2(t)}  \right)\tau =0.
    \label{eq8_app2}
\end{equation}
The phase modulation term $\phi_m$ is expressed using $\Omega_s$ as
\begin{eqnarray}
    \phi_m &=& \phi_{m0}\cos\left(\frac{2\Omega_s}{\sqrt{1-\frac{1}{8}\omega^2\Xi^2(t)}}t + \Delta\omega t \right) \nonumber \\
    &\approx& \phi_{m0}\cos \left( 2\Omega_s + \Delta\omega+\frac{1}{8}\Omega_s\omega^2\Xi^2(t)\ \right)t.
    \label{eq9_app2}
\end{eqnarray}
If we introduce the following values
\begin{equation}
    \epsilon' = \frac{\phi_{m0}}{1-\frac{1}{8}\omega^2\Xi^2(t)} \cot\phi_0,
    \label{eq10_app2}
\end{equation}
the equation of motion is finally written as
\begin{widetext}
\begin{equation}
    \frac{d^2\tau}{dt^2}+2\gamma_\epsilon\frac{d\tau}{dt}
    +\Omega_s^2\left(1+\epsilon' \cos\left(2\Omega_s+\Delta\omega+\frac{1}{8}\Omega_s\omega^2\Xi^2(t)\right)t\right)\tau =0,
    \label{eq11_app2}
\end{equation}
\end{widetext}
which corresponds to Eq. (\ref{eq22_sec2}).

\nocite{*}
\bibliography{RFPMmochi_ref1.bib}

\end{document}